\documentclass[sigconf]{acmart}

\renewcommand\footnotetextcopyrightpermission[1]{} 
\makeatletter
\renewcommand\@formatdoi[1]{\ignorespaces}
\makeatother

\newif\ifediting
\editingtrue

\newif\ifreindent

\usepackage{graphicx}
\usepackage{subfig}

\graphicspath{{./images/}} 

\usepackage{color} 

\usepackage{multirow}

\def\FirstMonthExtensions{12,164}

\def\LastMonthExtensions{12,497}

\def\stableExtensions{8,810}
\def\percentageStableExtensions{72.4}

\def\detectedExtensions{1,110}
  
\def\AllExtensions{16,743}

\def\AverageExtensions{13k}

\def\PercentageExtensions{28}
\def\UsedExtensions{\detectedExtensions}

\def\Logins{60}
\def\LoginsCSP{16}
\def\LoginsRedirect{44}

\def\InitialExperiments{22,904}
\def\InitialUsers{19,814}

\def\FinalExperiments{16,393}
\def\FinalUsers{16,393}
\def\chromeUsers{7,643}

\def\DextSize{5,474}
\def\DlogSize{9,492}
\def\DextAlogSize{3,919}
\def\DextOlogSize{11,047}

\def\ExtUnique{18.38}
\def\LogUnique{11.30}
\def\ExtOLogUnique{34.51}
\def\ExtNonePercent{66.61}
\def\LogNonePercent{42.1}
\def\ExtOLogNonePercent{32.61}

\def\DextUnique{54.86\%}
\def\DextStableUnique{50.35\%}
\def\DlogUnique{19.53\%}
\def\DextAlogUnique{89.23\%}
\def\DextOlogUnique{51.15\%}

\def\DextUniqueE{39.29\%}

\def\DextAtleastTwoUnique{76.25\%}
\def\DextAtleastThreeUnique{92.22\%}
\def\DextAtleastFourUnique{95.85\%}

\def\DlogAtleastFiveUnique{31.58\%}

\def\DlogAtleastEightUnique{38.98\%}

\def\DextOneUnique{7.39\%}
\def\DextTwoUnique{45.35\%}
\def\DextThreeUnique{85.89\%}
\def\DextFourUnique{93.87\%}

\def\DlogOneUnique{0.10\%}
\def\DlogTwoUnique{7.82\%}

\def\NoJSUnique{22.98\%}

\def\MaxExtPerUser{33}
\def\MaxLogPerUser{40}

\def\usenixLikeUnique{55.64\%}
\def\xhoundLikeUnique{49.60\%}

\def\ThresholdNumerousExperiments{more than 4}
\def\UsersMoreThanThreeExperiments{66}

\def\UsersWithEmptyRecords{2,015} 
\def\ChromeUsersDetectionError{6}
\def\NonChromeUsersDetectionSuccess{261}
\def\MobileUsers{1,042}
\def\BraveBrowser{31} 

\def\NELogins{0.441} %
\def\NEExtensions{0.641}
\def\NEUserAgent{0.474}
\def\NECanvas{0.611} %
\def\NEPlugins{0.343}
\def\NETimezone{0.168}
\def\NEScreen{0.271}
\def\NEFonts{0.652}

\usepackage[utf8]{inputenc}
 
\def\code#1{\texttt {{#1}}}

\usepackage[utf8]{inputenc}
\usepackage{color}
\definecolor{lightgray}{rgb}{0.95, 0.95, 0.95}
\definecolor{darkgray}{rgb}{0.4, 0.4, 0.4}
\definecolor{purple}{rgb}{0.65, 0.12, 0.82}
\definecolor{editorGray}{rgb}{0.98, 0.98, 0.98}
\definecolor{editorOcher}{rgb}{0.2, 0.4, 0.6} 
\definecolor{editorGreen}{rgb}{0, 0.5, 0} 
\usepackage{listings}
\usepackage{hyperref}
\usepackage{amsmath}
\usepackage{amssymb}
\usepackage{multirow}

\ifediting
\let\mylabel\label
\def\label#1{\mylabel{#1}\fbox{#1}}
\fi

\def\SHORTEN{\vspace*{-0.4cm}}

\def\Dext{\ensuremath{D_{\mathit{Ext}}}}
\def\Dlog{\ensuremath{D_{\mathit{Log}}}}
\def\DextAlog{\ensuremath{\Dext \cap \Dlog}}
\def\DextOlog{\ensuremath{\Dext \cup \Dlog}}

\newif\iffull

\fullfalse 

\author{G\'abor Gy\"orgy Guly\'as}
\affiliation{INRIA}
\email{gabor.gulyas@inria.fr}

\author{Doli\`ere Francis Som\'e}
\affiliation{INRIA}
\email{doliere.some@inria.fr}

\author{Nataliia Bielova}
\affiliation{INRIA}
\email{nataliia.bielova@inria.fr}

\author{Claude Castelluccia}
\affiliation{INRIA}
\email{claude.castelluccia@inria.fr}
  
\renewcommand{\shortauthors}{G. Gy. Guly\'as et al.}  

\begin{document}

\title[On the Uniqueness of Browser Extensions and Web Logins]{To Extend or not to Extend: on the Uniqueness of Browser Extensions and Web Logins}

  \begin{abstract}
Recent works showed that websites can detect browser extensions 
 that users install and websites they are logged into. This poses significant
 privacy risks, since extensions and Web logins 
 that reflect user's behavior, can be used to uniquely identify users 
on the Web. \\
This paper reports on the first large-scale behavioral uniqueness study 
based on \FinalUsers\ 
users who visited our website.
We test and detect the presence of \AllExtensions\ Chrome extensions, 
covering 28\% of all free Chrome extensions. 
We also detect whether the user is connected to \Logins\ different websites.\\
We analyze how unique users are based on their behavior, 
and find out that 
 \DextUnique\ of users that have installed at least one detectable extension are unique; 
 \DlogUnique\ of users are unique 
 among those who have 
 logged into one or more detectable websites; 
 and \DextAlogUnique\ are unique 
 among users with at least 
 one extension and one login.\\ 
%
We use an advanced fingerprinting algorithm and show that it is possible to 
identify a user in less than 625 milliseconds by selecting the most 
unique combinations of extensions. \\
Because privacy extensions contribute to the uniqueness of users, 
we study the trade-off between the amount of trackers blocked by 
such extensions and how unique the users of these extensions are.
We have found that 
privacy extensions should be considered more useful than harmful.
The paper concludes with possible countermeasures.

\end{abstract}

\maketitle

\section{Introduction}

\def\question#1{\textbf{#1}}

In the last decades, researchers 
have been actively studying users' uniqueness in 
various fields, in particular biometrics and privacy communities hand-in-hand 
analyze various characteristics 
of people, their behavior and the systems they are using. 
Related research showed that a person can be characterized based on her 
typing behavior~\cite{Roth-etal-14-IEEETIP,Zhon-etal-12-CVPRW}, 
mouse dynamics~\cite{Pusa-Brod-04-VDMCS}, 
and interaction with websites~\cite{Gamb-etal-07-BS}.
Furthermore, Internet and mobile devices provide rich environment where users'
habits and preferences can be automatically detected.
Prior works showed that users can be uniquely identified based on 
websites they visit~\cite{Olej-Cast-12-HotPets}, 
 smartphone apps they install~\cite{Jagd-Acs-Cast-15-WPES}
 and mobile traces they leave behind them~\cite{Mont-etal-13-SR}. 

Since the web 
browser is the tool people use to navigate through the Web, 
privacy research community has 
studied various forms of 
\emph{browser fingerprinting}~\cite{Ecke-10-PETS,Acar-etal-13-CCS,
Niki-etal-13-SP,Engl-Nara-16-CCS,Cao-etal-17-NDSS,Gome-etal-18-WWW}. 
Researchers have shown that a user's browser has a number of ``physical'' characteristics 
that can be used to uniquely identify her browser and hence to track it across 
the Web.
Fingerprinting of users' devices is similar to physical biometric traits of people, 
where only physical characteristics are studied. 

Similar to previous demonstrations of user uniqueness based on 
their 
behavior~\cite{Olej-Cast-12-HotPets,Jagd-Acs-Cast-15-WPES}, 
\emph{behavioral characteristics}, such as 
browser settings and the way people use their browsers can also help 
to uniquely identify Web 
users. 
For example, a user installs web browser extensions she prefers, such as 
AdBlock~\cite{AdBlock}, LastPass~\cite{LastPass} or Ghostery~\cite{Ghostery}
 to enrich her Web experience. Also, while browsing the Web, she
 logs into her favorite social networks, such as Gmail~\cite{Gmail}, 
 Facebook~\cite{Facebook} or LinkedIn~\cite{LinkedIn}.
In this work, we study \emph{users' uniqueness} based on their behavior and preferences on the Web:
we analyze how unique are Web users based on their \emph{browser extensions and logins}.

In recent works, Sj\"{o}sten et al.~\cite{Sjos-etal-17-CODASPY} and 
Starov and Nikiforakis~\cite{Star-Niki-17-IEEESP} explored two complementary 
techniques to detect extensions.
S{\'{a}}nchez{-}Rola et al.~\cite{Sanc-etal-17-USENIX} then discovered how to detect 
 any extension via a timing side channel attack. These works were focused on the technical 
mechanisms to detect extensions, 
but what was not studied is \emph{how browser extensions contribute 
to uniqueness of users at large scale}. 
Linus~\cite{Linus16}  showed 
that some social websites are vulnerable to the ``login-leak'' attack
that allows an arbitrary script to detect whether a user  is logged into a vulnerable website. 
However, it was not studied \emph{whether Web logins can also contribute to users' uniqueness}.

In this work,
we performed the first large-scale study of user uniqueness
based on  browser extensions and Web logins, 
collected from more than 16,000 users who visited our website (see the breakdown in Fig. \ref{fig:dataset_subsets}). 
Our experimental 
website identifies installed Google Chrome~\cite{GoogleChrome} extensions via Web Accessible Resources~\cite{Sjos-etal-17-CODASPY}, 
and detects 
websites where the user is logged into,
by methods that rely on URL redirection and CSP violation reports. 
Our website is able 
to detect the presence of \AverageExtensions\ Chrome extensions on average per month (the number of detected 
extensions varied monthly between $12,164$ and $13,931$), 
covering approximately $28\%$ of all free Chrome extensions\footnote{The list of detected extensions and websites are available on our website: \url{https://extensions.inrialpes.fr/faq.php}}
. 
We also detect 
whether the 
user is connected to one or more of \Logins\ different websites.
Our main contributions are:
\begin{itemize}
\item A large scale study on \emph{how unique users are based on their browser extensions and website logins}.
We discovered that \DextUnique\ 
of users that have installed at least one detectable extension are unique; 
 \DlogUnique\ of users are unique among those who have logged into one or more detectable websites; 
 and \DextAlogUnique\ are unique among users with at least one extension and one login.
Moreover, we discover that \NoJSUnique\ of users could be uniquely identified 
by Web logins, even if they disable JavaScript.
\item We study the privacy dilemma on Adblock and privacy extensions, that is, \emph{how well
these extensions protect their users against trackers and how they also contribute 
to uniqueness}. We evaluate the statement ``the more privacy extensions you 
install, the more unique you are'' by analyzing how users' uniqueness increases with 
the number of privacy extensions they install; and by evaluating the 
tradeoff between the privacy gain of the blocking extensions
such as Ghostery~\cite{Ghostery} and Privacy Badger~\cite{PrivacyBadger}. 
\end{itemize}

\begin{figure}[!t]
\centering
\includegraphics[width=0.35\textwidth]{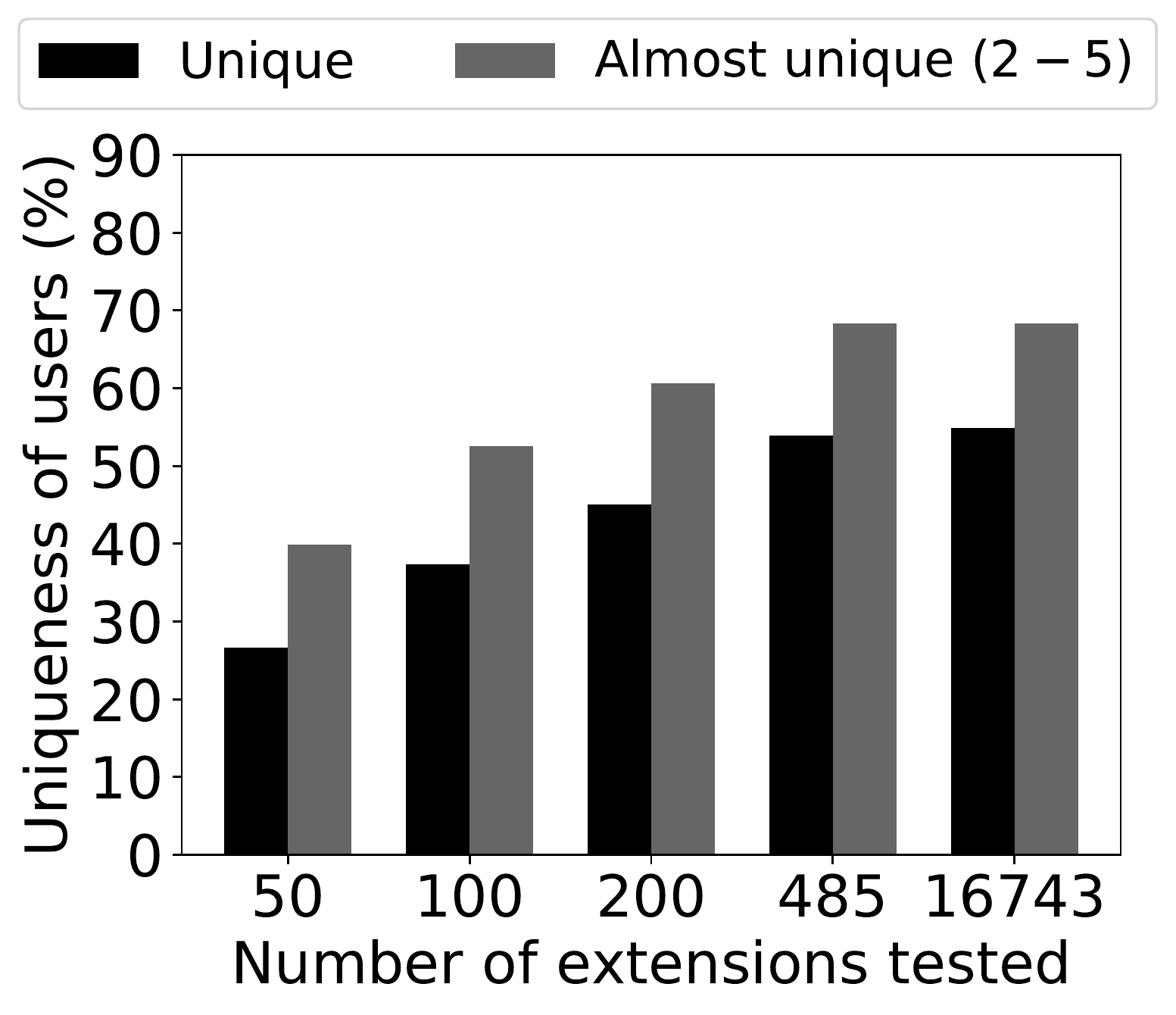}
\caption{Results of general fingerprinting algorithm. 
{\normalfont Testing  485 carefully selected extensions 
provides a very similar uniqueness result 
to testing all 16,743 extensions. Almost unique means that there are 2--5 users with the same fingerprint.}}
\label{fig:uniqueness_vs_fingerprinting}
\SHORTEN
\end{figure}

We furthermore show
that browser extensions and Web logins
can be exploited to fingerprint and track users by only checking a limited
number of extensions and Web logins.
We have applied an advanced fingerprinting algorithm~\cite{Guly-Acs-Cast-16-PETS} 
that carefully selects a limited number of extensions and logins. 
For example, 
Figure~\ref{fig:uniqueness_vs_fingerprinting}  shows the uniqueness of 
users we achieve by testing a limited number of extensions. The last column 
shows that \DextUnique\ of users are unique based on all 
\AllExtensions\ detectable extensions. 
However, by testing 485 carefully chosen extensions we can identify more than 53.96\% of users. 
Besides, detecting 485 extensions takes only 625ms.

Finally, we give suggestions to the end users as well as 
website owners and browser vendors on 
how to protect 
the users from the fingerprinting based on extensions and logins.

In our study we did not have enough data to make any claims about the 
stability of the browser extensions and web logins because only few users 
repeated an experiment on our 
website (to be precise, only \UsersMoreThanThreeExperiments\ users out of \FinalUsers\ users 
have made \ThresholdNumerousExperiments\ tests on our website). We leave this as a future work.

\section{Background}
\label{sec:background}


\begin{figure*}[!t]
\centering
\includegraphics[width=6.7in]{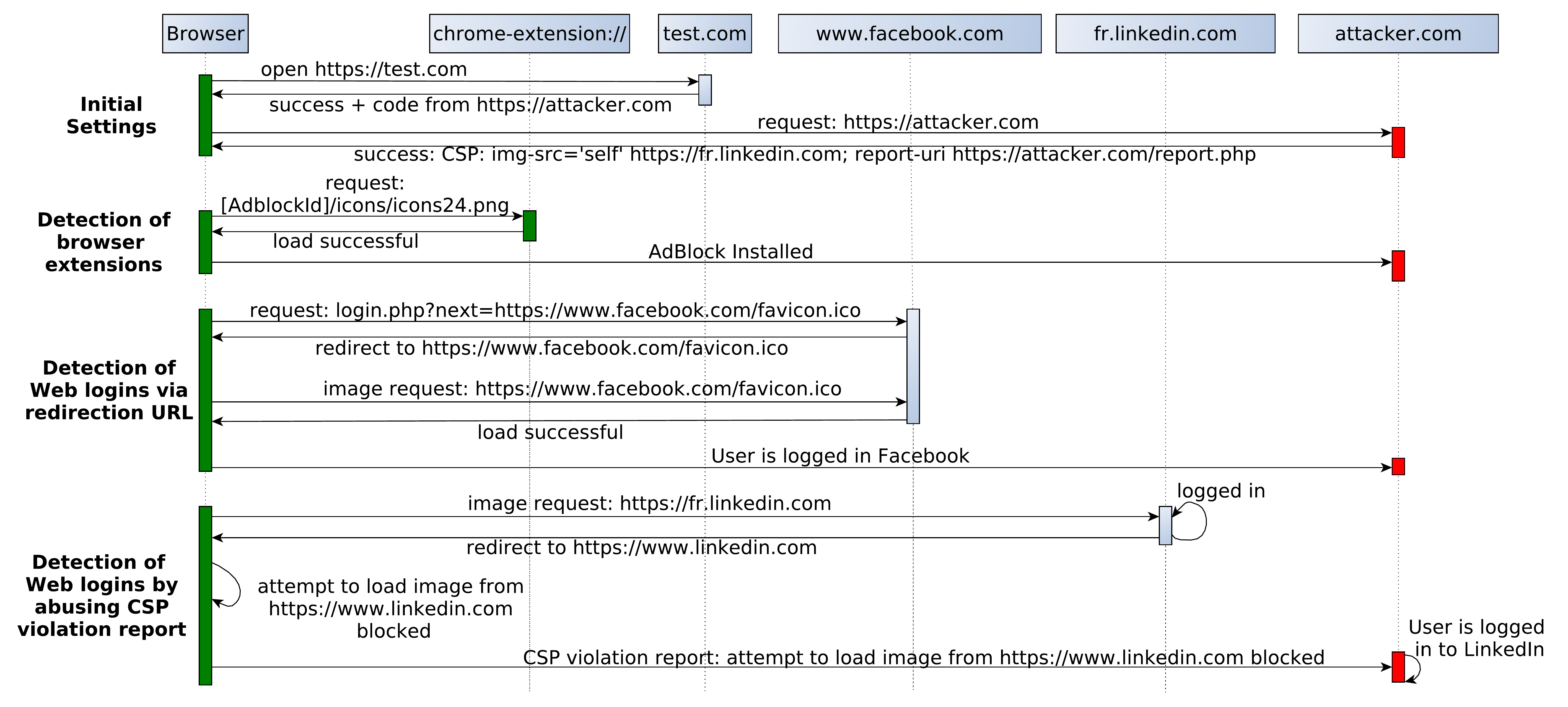}
\caption{Detection of browser extensions and Web logins. {\normalfont A user visits a benign website \url{test.com} which embeds third party code (the attacker' script) from \url{attacker.com}. The script detects an icon of \code{Adblock} extension and concludes that \code{Adblock} is installed. Then the script detects that the user is logged into Facebook when it successfully loads Facebook \code{favicon.ico}. It also detects that the user is logged into LinkdedIn through a CSP violation report triggered because of a redirection from \url{https://fr.linkedin.com} to \url{https://www.linkedin.com}. All the detection of extensions and logins are invisible to the  user.}}
\label{fig:redirection}
\end{figure*}

\subsection{Detection of browser extensions}

A \emph{browser extension} is a program, typically written in JavaScript, HTML and CSS.  
An extension may alter the context of the webpage, 
intercept HTTP requests, 
or provide extra functionality. 
Examples of browser extensions are ad blockers, such as AdBlock~\cite{AdBlock} and  password 
managers, such as LastPass~\cite{LastPass}. 

\def\WARs{\emph{web\_accessible\_resources}}

In the Google Chrome web browser, each extension comes with 
\emph{a manifest file}~\cite{ManifestFile}, which contains metadata 
about the extension.
Each extension has a unique and permanent identifier, and the manifest file
of an extension with identifier \code{extID} is located at 
\code{chrome-extension://[extID]/manifest.json}.
The manifest file has a section \WARs\ (WARs) that 
declares which resources of an extension are accessible in the content of any webpage~\cite{WARs}. 
The WARs section specifies a list of paths to such resources,
presented by the following type of URL: \\
 \code{chrome-extension://[extID]/[path]},
 where \code{path} is 
the path to the resource in the extension.

Therefore, a script that 
tries to load such 
an accessible resource 
in the context of an arbitrary webpage 
is able to 
check whether an extension is installed with a 100\% guarantee:
if the resource is loaded, an extension is installed, otherwise it is not.
Figure~\ref{fig:redirection} shows an example of AdBlock extension detection: 
the script tries to 
load an image, which is declared in the \WARs\ section of AdBlock's manifest file. 
If the image from AdBlock, located at \code{chrome-extension://[AdBlockID]/icons/icons24.png} 
is successfully loaded, 
then AdBlock is installed in the user's browser.

Sj\"{o}sten et al.~\cite{Sjos-etal-17-CODASPY} were the first to crawl the Google Chrome Web 
Store and to discover that 28\% of all free Chrome extensions are detectable by WARs.
An alternative method to detect extensions that was available at the beginning of our experiment, 
was a behavioral method from XHOUND~\cite{Star-Niki-17-IEEESP}, but it 
had a number of false positives and detected only 9.2\% of top 10k extensions 
(see Sec.~\ref{sec:related} for more details).  
Therefore, we decided to reuse the code from Sj\"{o}sten et 
al.~\cite{Sjos-etal-17-CODASPY} with their permission to crawl Chrome Web 
Store and identify detectable extensions based on WARs.
During our experiment, we discovered that WARs could be detected in other 
Chromium-based browsers like Opera~\cite{OperaBrowser} and the Brave Browser~\cite{BraveBrowser} 
(we could even detect Brave Browser since 
it is shipped with several default extensions
detectable by WARs). 
We have chosen to work with Chrome, as it was the most affected.

\subsection{Detection of Web logins}
In general, 
a website cannot detect whether a user is logged into other websites because of 
Web browser security mechanisms, such as access control 
and Same-Origin Policy~\cite{Same-Origin-Policy}. 
In this section, we present two advanced methods that, despite browser security mechanisms, 
allow an attacker to detect the websites where the user is logged into. 
Figure~\ref{fig:redirection} presents all the detection mechanisms. 

\noindent {\bf Redirection URL hijacking.}
The first requirement for this method to work is the login redirection mechanism: 
when a user is not logged into Facebook, and tries to access 
an internal Facebook resource, she automatically gets redirected to the URL 
\url{http://www.facebook.com/login.php?next=[path]}, 
where \code{path} is the path to the resource.
The second requirement is that the website should have an internal 
image available to all the users. In the case of Facebook, it is a \code{favicon.ico} image.

By dynamically embedding an image pointing to 
\code{\url{https://www.facebook.com/login.php?next=https\%3A\%2F\%2Fwww.facebook.com\%2Ffavicon.ico}}
into the webpage, an attacker can detect whether the user is logged into Facebook or not. 
If the image loads, then the user is logged into Facebook, otherwise she is not.
This method has been shown to successfully 
detect logins on dozens of websites~\cite{Linus16}.

\noindent {\bf Abusing CSP violation reporting.}
Content-Security-Policy (CSP)~\cite{Stam-Ster-Mark-10-WWW,CSP-MDN}
is a security mechanism that allows programmers to control which 
client-side resources can be loaded and executed by the browser. 
CSP (version 2) is an official W3C candidate recommendation~\cite{CSP2-W3C}, 
and is currently supported by major web browsers.

A CSP delivered with a page controls the resources of the page.
For example, CSP can set a rule to allow the browser to load images
only from a particular domain. If a webpage tries to load an image 
from a different domain, CSP will block such request and can send 
a violation report back to the web server.
%

An attacker can misuse CSP  to detect redirections~\cite{Homakov14}. 
We extend this idea to detect logins. 
For this method to work, a website should redirect its logged in 
users to a different domain. In the case of LinkedIn, 
the users, who are not logged in, visit   \code{fr.linkedin.com}, 
while the users, who are logged in, are automatically redirected to 
a different domain \code{www.linkedin.com}.
The lowest block of 
Fig.~\ref{fig:redirection} presents an example of such attack on LinkedIn.
Initially, the attacker embeds a hidden iframe from his own domain 
with the CSP that restricts loading images only from \code{fr.linkedin.com}. 
Then, the attacker dynamically embeds a new image on the testing website, 
pointing to \code{fr.linkedin.com}. If the user is logged in, LinkedIn will 
redirect her to the \code{www.linkedin.com}, and thus the browser 
will fire a CSP violation report because images can be loaded only 
from \code{fr.linkedin.com}. By receiving the CSP report, the attacker 
deduces that the user is logged in LinkedIn.


\section{Dataset}
\label{sec:dataset}

We launched an experiment website in April 2017
to collect browser extensions and Web logins with the
goal of studying users' uniqueness at a large scale.
We have advertised our experiment by all possible means,
including social media and in press.
In this section, we first 
present the set of attributes that we collect
in our experiment 
and
the rules we applied to filter out irrelevant records.
Then, we provide data statistics and show which
extensions and logins are popular among our users. 

\subsection{Experiment website and data collection}
\label{sec:experiment}

\begin{figure}[!ht]
\centering
\includegraphics[width=0.45\textwidth]{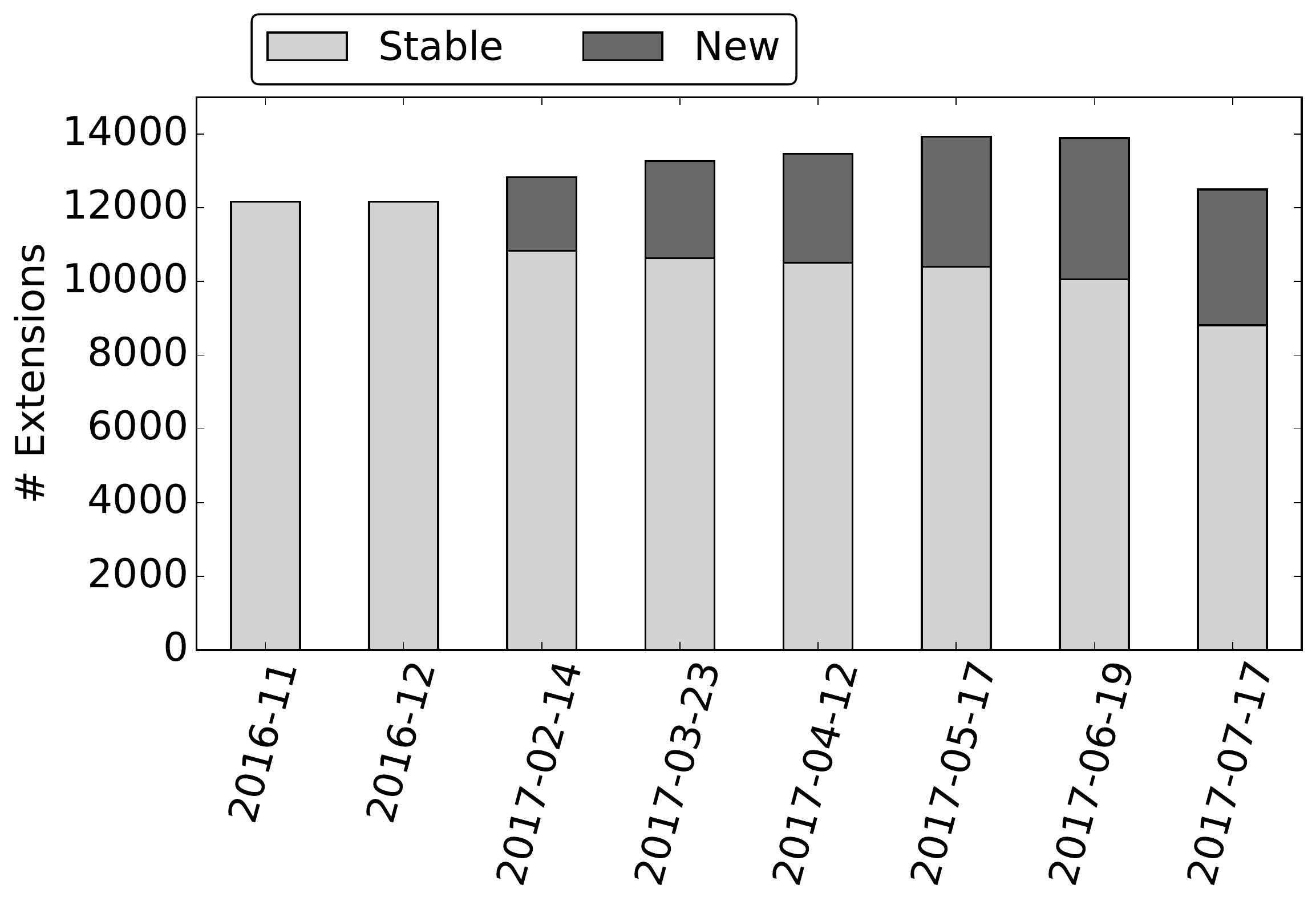}
\caption{Evolution of detected extensions in Chrome}
\label{fig:evolution_extensions_wrt_first_month}
\end{figure}

The goal of our  website is both to collect
browser extensions and Web logins, and to inform users
about privacy implications of this particular type of
fingerprinting. Using the various detection techniques described in
Section~\ref{sec:background}, we collect the
following attributes:
\begin{itemize}
\item The list of installed browser extensions, using 
web accessible resources. 
For each user 
we tested 
around 
\AverageExtensions\ extensions detectable at the moment of testing (see Figure~\ref{fig:evolution_extensions_wrt_first_month}).
%
\item The list of Web logins: we test for \LoginsRedirect\
logins using redirection URL hijacking and \LoginsCSP\
logins using CSP violation report.
\item Standard fingerprinting attributes~\cite{Lape-etal-16-SP}, 
such as fonts installed, Canvas fingerprint~\cite{Acar-etal-14-EEJND}, 
and WebGL~\cite{Mowe-Shac-12-W2SP}. 
To collect these attributes, we use FingerprintJS2, 
which is an open-source browser fingerprinting library~\cite{FingerprintJS2}.
We collected these attributes in order to clean our data and 
compare entropy with other studies (see Table~\ref{tab:widgets}).
\end{itemize}

To recognize users that perform several tests on our website, we have stored 
a unique identifier for each user in the HTML5 localStorage. 
%
We have communicated our website via 
forums and social media 
channels related to science and technology, and got press coverage in 3 newspapers. 
We have collected  \InitialExperiments\ experiments performed by \InitialUsers\ users between April and August 2017. 

{\bf Ethical concerns.}
Our study was validated by an IRB-equivalent service at our institution. All visitors are informed of our goal, and are provided with both Privacy Policy and FAQ sections of the website. The visitors have to explicitly click on a button to trigger the collection of their browser attributes. In our Privacy Policy, we explain what data we are collecting, and give a possibility to opt-out of our experiment. The data collected is used only for our own research, will be held until December 2019 and will not be shared with anyone.

%
%
%
\begin{table}[!ht]
\caption{Users filtered out of the final dataset}
\begin{tabular}{p{6.7cm}r}
\hline
Initial users & \bf{\InitialUsers} \\ \hline
Mobile browser users & \MobileUsers \\ \hline
Chrome browser users with extension detection error & \ChromeUsersDetectionError \\ \hline
Non Chrome users with at least one extension detected & \NonChromeUsersDetectionSuccess \\ \hline
Brave browser users & \BraveBrowser \\ \hline
Users whose browser has an empty user-agent string, screen resolution, fonts, or canvas fingerprint &  \UsersWithEmptyRecords \\ \hline
Users with \ThresholdNumerousExperiments\ experiments & \UsersMoreThanThreeExperiments \\ \hline
Final dataset & \bf{\FinalUsers} \\ \hline
Chrome browser users in the final dataset & \chromeUsers \\
\hline
\end{tabular}
\label{tab:cleaning_rules}
\end{table}

{\bf Data cleaning.} 
We applied a set of cleaning rules over our initial data, to improve the quality of the data. 
The final dataset contains \FinalExperiments\ valid experiments (one per user). 
Table~\ref{tab:cleaning_rules} shows the initial number of users and which users have been removed from our initial dataset. 
We have removed all \MobileUsers\ users with mobile browsers. At the time of writing this paper, browser 
extensions were not supported on Chrome for mobiles. Since extensions detection were designed 
for Chrome, we then excluded mobile browsers. Moreover, mobile users tend to prefer native apps 
rather than their web versions\footnote{\url{https://jmango360.com/wiki/mobile-app-vs-mobile-website-statistics/}}. 
In fact, the popular logins 
in our dataset, such as  Gmail, Facebook, Youtube, 
all have a native mobile version. 

We have also removed \UsersWithEmptyRecords\ users that have deliberately tampered with their browsers: 
for example, users with empty user-agent string, empty screen resolution or canvas fingerprint. 
We think that it is reasonable not to trust information received from those users, as they may have tampered with it. 
We also needed this information to compare our study with previous works on browser fingerprinting.
Finally, we have excluded  users who have tampered with extension detection. This includes Chrome users 
for whom extension detection did not successfully complete, and users of 
other browsers with at least 1 extension detected. 

For users who visited our website and performed up to 4 experiments, we kept only one 
experiment, the one with the biggest number of extensions and logins. 
We then removed 
\UsersMoreThanThreeExperiments\ users with more than 4 experiments. 
We suspect that the goal of such users with numerous experiments 
was just to use our website in order to test the uniqueness of their browsers with different browser settings. 

{\bf Evolution of browser extensions.}
From November 2016 to July 2017, we crawled on a montly basis the 
free extensions on the Chrome Web Store
in order to keep an up-to-date set of extensions for our experiment. 
Figure~\ref{fig:evolution_extensions_wrt_first_month} presents the evolution of 
extensions throughout the period of our experiment. 
Since some extensions got removed from the Chrome Web Store, the number of stable extensions decreased. 


Out of \FirstMonthExtensions\ extensions that were detectable in November 2016, 
\stableExtensions\ extensions (\percentageStableExtensions\%) remained stable throughout 
the 9-months-long experiment. In total, 
\AllExtensions\ extensions were detected at some point 
during these 9 months. Since every month the number of detectable extensions was different, 
on average we have tested around \AverageExtensions\ extensions during each month.

\subsection{Data statistics}
\label{sec:datastats}

Our study is the first to analyze 
uniqueness of users based on their browser extensions and logins at large scale. 
Only uniqueness based on browser extensions was previously measured, 
but on very small datasets of 204
~\cite{Sanc-etal-17-USENIX} and  854
~\cite{Star-Niki-17-IEEESP} participants. 
We measure uniqueness of \FinalUsers\ users for all attributes, and 
of  \chromeUsers\ Chrome browser users for browser extensions.

{\bf Comparison to previous studies.} 
To compare our findings with the previous works on browser extensions, 
we randomly pick 
subsets of 204 (as in \cite{Sanc-etal-17-USENIX}) and 854 (as in \cite{Star-Niki-17-IEEESP}) 
Chrome users 100 times (we found that 
picking 100 times provided a stable result). 
Table~\ref{tab:previous_studies} shows uniqueness results from previous works
and an estimated uniqueness using our dataset.

\begin{table}[!t]
\caption{\label{tab:previous_studies}Previous studies on measuring uniqueness based on browser extensions and our estimation of uniqueness.}
\centering
\begin{tabular}{p{1.2cm}p{1.4cm}p{1.3cm}p{1.2cm}p{1.5cm}}
\hline
{Study } 	& Fingerprints collected in a study &  Extensions targeted in a study& Unique fingerprints  in a study & Unique fingerprints in our dataset \\
\hline
\hline
Timing leaks~\cite{Sanc-etal-17-USENIX}  & 204 & 2,000 &  56.86\% &  \usenixLikeUnique\ \\ 
XHOUND \cite{Star-Niki-17-IEEESP}  	& 854 & 1,656 & 14.10\% & \xhoundLikeUnique\ \\
\hline
Ours & \chromeUsers\ & \AverageExtensions\ & \DextUniqueE & \DextUniqueE \\
\hline
\end{tabular}
\end{table}

The last column in Table~\ref{tab:previous_studies} 
shows our evaluation of uniqueness for a given subset of users.
Our estimation for 204 random users is  \usenixLikeUnique, which is close 
to the 56.86\%\ from the original study~\cite{Sanc-etal-17-USENIX}.
For 854 random users, 
we estimate that \xhoundLikeUnique\ of them are unique, 
while in the original XHOUND study~\cite{Star-Niki-17-IEEESP}
the percentage of unique users is only  14.1\%.
We think that such small percentage of unique users in~\cite{Star-Niki-17-IEEESP} is due to (1) 
a smaller number of extensions detected (only 174 extensions were detected for 854 users); 
(2) 
a different user base: while our experiments and ~\cite{Sanc-etal-17-USENIX} 
targeted colleagues, students and other likely privacy-aware experts, 
XHOUND~\cite{Star-Niki-17-IEEESP} used Amazon Mechanical Turk, 
where users probably have different habits to installing extensions.   
Out of \chromeUsers\ users of the Chrome browser, where we detected 
extensions, \DextUniqueE\ of users were unique. This number shows a more 
realistic estimation of users' uniqueness based on browser extensions 
than previous works because of a significantly larger dataset. 

To the best of our knowledge, our study is the first to \emph{analyze uniqueness of 
users based on their web logins}, and on combination of extensions and logins.

{\bf Normalized Shannon's entropy.}
We compare our dataset with 
the previous studies on browser fingerprinting:  AmIUnique~\cite[Table B.3]{Lape-17-PhDThesis} 
(contains 
390,410 fingerprints, 
collected between November 2014 and June 2017) 
and Hiding in the Crowd~\cite{Gome-etal-18-WWW} (contains 1,816,776 users collected in 2017). 
Entropy measures the amount of identifying information in a fingerprint -- the higher 
the entropy is, the more unique and identifiable a fingerprint will be. To compare 
with previous datasets, which are of different sizes, we compute normalized Shannon's entropy:

\begin{equation}
H_N(X) = \frac{H(X)}{\log_2 N} = -\frac{1}{\log_2 N} \cdot \sum_{i} P(x_i) \log_2 P(x_i)
\end{equation}
where $X$ is a discrete random variable with possible values $\{ x_1, ..., x_n\}$, $P(X)$ is a 
probability mass function and $N$ is the size of the dataset.

Table~\ref{tab:fp-entropies} compares the entropy values of  well-known attributes for standard fingerprinting 
and for 
logins for all \FinalUsers\ users in our dataset 
and 
for browser extensions for \chromeUsers\ Chrome users.
All the attributes in standard fingerprinting are similar to previous works, except for fonts and plugins. 
Unsurprisingly, plugins entropy is very small because of decreasing support of plugins in Firefox~\cite{FirefoxNPAPI} and Chrome~\cite{ChromeNPAPI}.  
Differently from previous studies that detected fonts with Flash, we used JavaScript based font detection, relying on a list of 500 fonts shipped along with the FingerprintJS2 library. As those fonts are selected for fingerprinting, this could explain why our list of fonts provides a very high entropy.

\begin{table}[!t]
\caption{\label{tab:widgets}Normalized entropy of extensions and logins compared to 
previous studies.}
\label{tab:fp-entropies}
\centering
\begin{tabular}{p{2.5cm} p{1.2cm} p{1.2cm} p{1.8cm} p{1.5cm} }
\hline
\multicolumn{4}{c}{Standard fingerprinting studies} \\
\hline \hline
{\em Attribute} & {Ours} &  AmIUnique \cite{Lape-17-PhDThesis}  & Hiding~\cite{Gome-etal-18-WWW} Desktop  \\ \hline
User Agent & \NEUserAgent &   0.601 & 0.304 \\
List of Plugins & \NEPlugins &   0.523  & 0.494\\
Timezone & \NETimezone &  0.187 & 0.005 \\
Screen Resolution & \NEScreen &  0.276 & 0.213 \\
List of Fonts  & \NEFonts &  0.370 & 0.335 \\
Canvas & \NECanvas & 0.503 & 0.387 \\
\hline
\hline
\multicolumn{4}{c}{Studies on extensions and logins} \\
\hline \hline
{\em Attribute} & Ours 
& Timing leaks~\cite{Sanc-etal-17-USENIX} & XHOUND \cite{Star-Niki-17-IEEESP} \\\hline
Extensions & \NEExtensions &  0.869 & 0.437 \\
Logins & \NELogins & N/A & N/A \\
\hline
\end{tabular}
\SHORTEN
\end{table}

In our dataset, as well as in previous studies, browser extensions are one of the most discriminating attributes of a user's browser.
The computed entropy of 0.641, computed for the \chromeUsers\ Chrome users, 
lays between the findings of Timing leaks~\cite{Sanc-etal-17-USENIX} and 
XHOUND~\cite{Star-Niki-17-IEEESP}. One possible explanation is the size of the user base.
For instance, users in XHOUND had few and probably often the same extensions detected 
(out of 1,656 targeted extensions, only 174 were detected for 854 users), making only 14.1\% of them
unique. This explains why the entropy in XHOUND is smaller.
S{\'{a}}nchez{-}Rola et al.~\cite{Sanc-etal-17-USENIX} computed a very high entropy, but on a very small dataset of 204 users: 116 of them had a unique set of installed extensions, and thus the computed entropy was very high.

\subsection{Usage of extensions and logins}




\begin{figure*}[!h]
   \begin {minipage}{0.96\textwidth}
     	\centering
     	\includegraphics[width=6.5in]{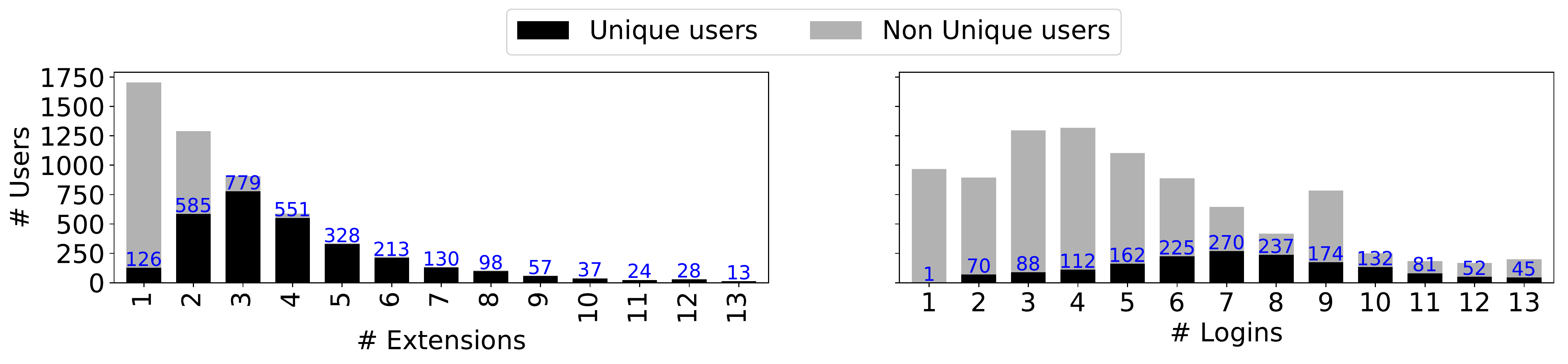}
   \end{minipage}
   \caption{Usage of browser extensions and logins by all users.}
   \label{fig:distribution_number_of_users}
\end{figure*}

Figure~\ref{fig:distribution_number_of_users} 
shows the distribution of users in our dataset according to the number of detected extensions 
and logins (users having between 1 and 13 logins or extensions detected), and the number of 
unique users as they are grouped by number of detected extensions and logins.
The maximum number of extensions we detected for a single user was \MaxExtPerUser. The 
number of users decreases with the number of 
extensions.
The largest group of users have only 1 extension detected, followed by users with 2 detected extensions, etc. 
%
We notice that the more extensions a user has, the more unique she is. We analyze this phenomenon further 
in Section~\ref{sec:uniqueness-final}.
%
Among users with exactly 1 extension detected, \DextOneUnique\ are unique.
This percentage rises to \DextTwoUnique\ and \DextThreeUnique\ 
for groups 
of users with exactly 2 and 3 detected extensions respectively.  

Figure~\ref{fig:distribution_number_of_users} also shows the distribution of users per number of detected logins. 
We found that most users have  between 1 and 10 logins, with a maximum number of  \MaxLogPerUser\ 
logins detected for one user. 
%
On our website, we were able to detect the presence of \Logins\ logins, which is rather small with respect to the 
large number of extensions we tested (around 13k per user). 
 %
 This explains why fewer users are unique based on their logins: for example, among users with exactly 1 
 login detected,    \DlogOneUnique\ are unique, and  \DlogTwoUnique\ are unique among users with 
 exactly  2 logins detected.

\begin{table}[!ht]
\caption{Top seven most popular extensions in our dataset and their popularity on Chrome Web Store}
\label{tab:popular_extensions}
\begin{tabular}{p{5cm}rl}
                          \bf{Extension} &  			\bf{Dataset} & 		\bf{Chrome} \\
                                 AdBlock &              1,557 &    10,000,000+ \\
         LastPass: Free Password Manager &              1,081 &      7,297,730 \\
                                Ghostery &               735 &       2,665,427 \\
                          Privacy Badger &               594 &         771,804 \\
                           Adobe Acrobat &               585 &     10,000,000+ \\
                   Cisco WebEx Extension &               482 &     10,000,000+ \\
                          Save to Pocket &               428 &       2,752,642 \\
\end{tabular}
\end{table}

\begin{table}[!ht]
\caption{Top seven most popular logins in our dataset and their ranking according to Alexa}
\label{tab:popular_logins}
\begin{tabular}{p{4.5cm}rr}
      \bf{Website} &  		 \bf{Dataset} &  \bf{Alexa Rank} \\
Gmail (subdomain of Google) &  6,828 &         1 \\
      Youtube &              6,780 &           2 \\
     Facebook &              5,493 &           3 \\
     LinkedIn &              3,913 &          13 \\
      Blogger &              3,393 &          53 \\
      Twitter &              3,274 &           8 \\
     eBay.com &              2,220 &          33 \\
\end{tabular}
\end{table}

{\bf What extensions are the most popular among our users?}
Table~\ref{tab:popular_extensions} presents the seven 
most detected extensions in our 
dataset of \FinalUsers\ users. 
The three most popular extensions 
are 
AdBlock~\cite{AdBlock}, password manager LastPass~\cite{LastPass} 
and tracker blocker Ghostery~\cite{Ghostery}. 
These extensions are also very popular according to their downloads statistics on Chrome Web Store.
%
%

{\bf What websites users are logging into the most?} Table~\ref{tab:popular_logins} shows the 
seven most detected websites in our experiment. These websites are also highly rated 
according to Alexa\footnote{Alexa ranking extracted on the the 28th of June 2018}. For instance, 
Google~\cite{Google}, Facebook~\cite{Facebook} and 
Youtube~\cite{Youtube} are regularly ranked 
as the top 3 most popular websites by Alexa\footnote{Note that Gmail is a subdomain of Google, that is why it is ranked 1 in Table~\ref{tab:popular_logins}.}. 
Being able to detect such popular websites further strengthen our study as they represent 
websites that are widely used by users in the wild.

\section{Uniqueness analysis}
\label{sec:uniqueness}

In this section we 
present the results for 
user's uniqueness 
based on all 
 \AllExtensions\ 
 extensions and \Logins\ logins.
We first show uniqueness for the full dataset of \FinalUsers\ users, 
and then present more specific results for various subsets of our dataset. 
 
\begin{figure}[!t]
   \begin{minipage}{0.48\textwidth}
     	\centering
    	\includegraphics[width=3.4in]{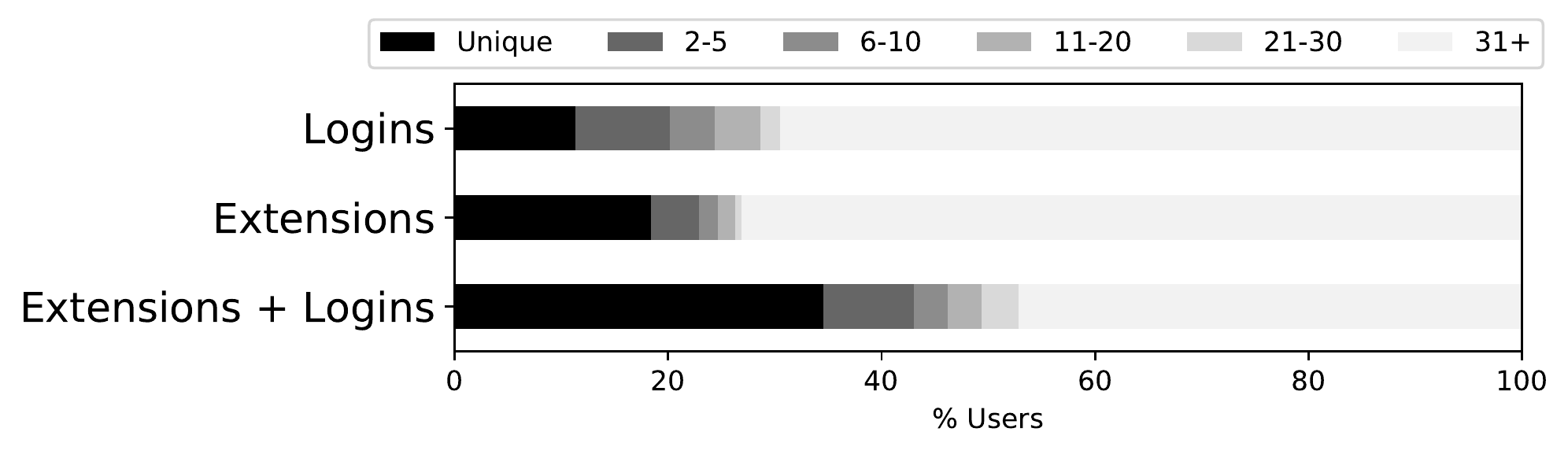}
   \end{minipage}
  \caption{Distribution of anonymity set sizes for \FinalUsers\ users based on detected extensions and logins.}
  \label{fig:anonymity_sets_full_dataset}
  \SHORTEN
\end{figure}

{\bf Uniqueness results for the full dataset.}
Figure~\ref{fig:anonymity_sets_full_dataset} shows the uniqueness of users according to their 
extensions and logins, and a combination of both attributes. Out of the \FinalUsers\ users, 
\LogUnique\% are unique based on their logins. 
For \LogNonePercent\% of users in 
our dataset, we did not detect any logins. These users either did not log into any of the \Logins\ 
websites we could detect or blocked third party cookies, that prevented our login detection from 
working properly. 

Considering only detected extensions, 
\ExtUnique\% of users in our dataset are unique. 
This result is also 
influenced by the \ExtNonePercent\% of users who did not have any extension 
detected: these are  either Chrome users with no extensions detected, or users of other browsers. 

An attacker willing to fingerprint users can also use their detected logins and extensions combined. 
Interestingly, by combining extensions and logins, we found that \ExtOLogUnique\% of users are uniquely identifiable. 
It is worth mentioning that \ExtOLogNonePercent\% of users have no extensions and no logins detected.
This impacts significantly the computed uniqueness.



\subsection{Four final datasets}
\label{sec:subsets}

\begin{figure}[!t]
\centering
\includegraphics[width=0.35\textwidth]{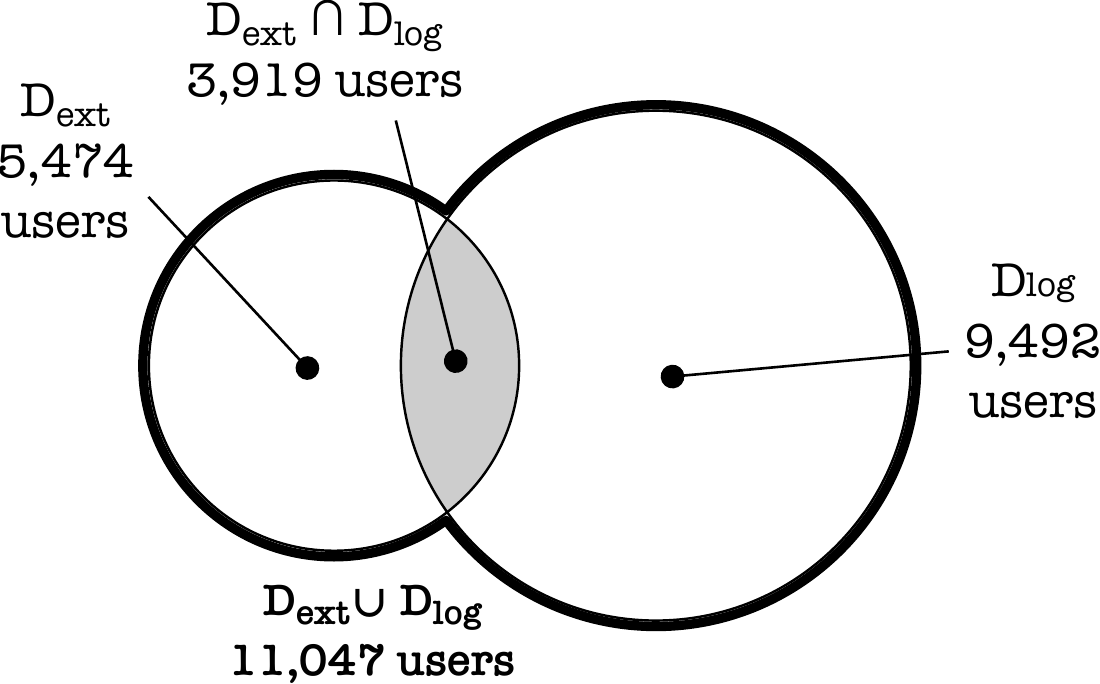}
\caption{Four final datasets.
{\normalfont \Dext\ contains users, who have installed at least one detected extension and 
\Dlog\ contains users, who have at least one login detected.}}
\label{fig:dataset_subsets}
\SHORTEN
\end{figure}

In our full dataset of  \FinalExperiments\ users, 
we have observed \chromeUsers\ users of Chrome browser, for whom  testing of 
browser extensions worked properly.
In this subsection we consider various subsets of our full dataset that demonstrate 
uniqueness results for users who have at least one extension or one login detected. 
Figure \ref{fig:dataset_subsets} shows four final datasets that we further analyze in this section:
\begin{itemize}
\item \Dext\ contains \DextSize\  Chrome users, who have installed at least one extension that we can detect. 
\item \Dlog\ contains \DlogSize\ users, who have logged into at least one website that we detect.
\item \DextAlog\ contains \DextAlogSize\ Chrome users who have at least one extension and one login detected. 
\item \DextOlog\ contains \DextOlogSize\ users who have either at least one extension or at least one login detected. 
\end{itemize} 


\subsection{Uniqueness results for final datasets}
\label{sec:uniqueness-final}

\begin{figure}[!t]
   \begin{minipage}{0.48\textwidth}
     	\centering
    	\includegraphics[width=3.4in]{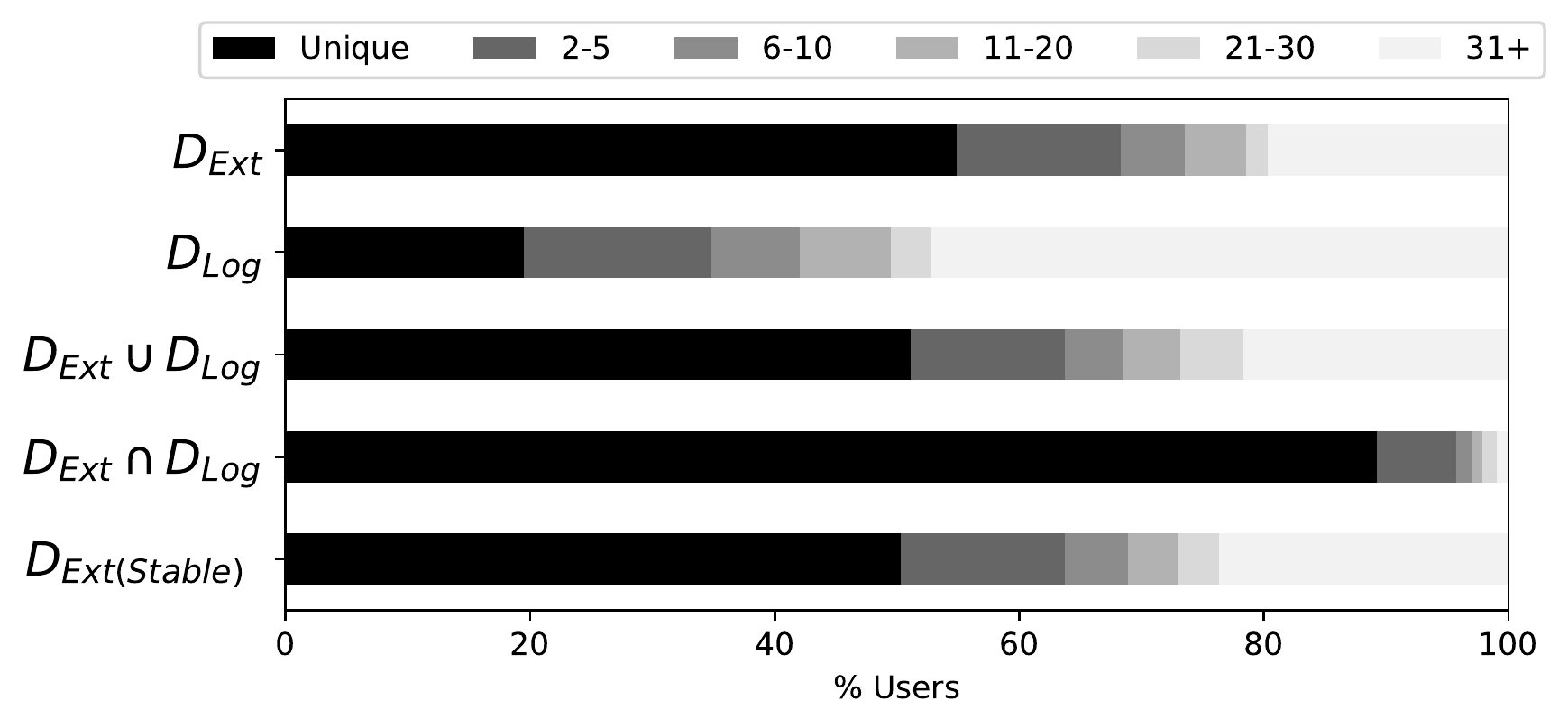}
   \end{minipage}
   \SHORTEN
  \caption{Anonymity sets for different datasets}
  \label{fig:anonymity_sets_datasets}
  \SHORTEN
\end{figure}

Figure~\ref{fig:anonymity_sets_datasets} presents results for the four datasets.
\Dext\ dataset shows that 
\DextUnique\ of users are uniquely identifiable among Chrome users,
who have at least one detectable extension.
This demonstrates that browser extensions detection is a serious privacy threat as a fingerprinting technique.

Among \DlogSize\ users with at least one login detected (\Dlog\ dataset) 
only \DlogUnique\ are uniquely identifiable. This result can be explained by 
a very small diversity of attributes (only \Logins\ websites).

When we analyzed Chrome users who have at least one extension and one login detected (\DextAlog\ dataset), 
we found out that  \DextAlogUnique\ of them are uniquely identifiable. 
This means that without any other fingerprinting attributes, the mere installation of at least one extension, in addition to being logged into at least one website imply that the majority of users in this dataset can be tracked by their fingerprint based solely 
on extensions and logins! 

Furthermore, for dataset \DextOlog\ that contains users with at least one extension or at least one login, 
we compute that \DextOlogUnique\ of users can be uniquely identified. 
This result becomes particularly interesting when we compare the size of the \DextOlog\ dataset, which contains \DextOlogSize\ users, 
with the size of the \Dext\ dataset, that has \DextSize\ users. The size of \DextOlog\ is almost twice as large as \Dext. 
%
Nevertheless, the percentage of unique users and the distribution of anonymity set sizes in these datasets are very similar:
\DextUnique\ of unique users in \Dext\ and \DextOlogUnique\ of unique users in  \DextOlog.
We believe this is due to the fact that extensions and logins are orthogonal properties. 
We checked the cosine similarity between these attributes as binary vectors, and found that all attribute pairs had a very low similarity score, all below $0.34$, with 11 exceptions below $0.2$.

%
%

The last row $D_{\mathit{Ext(Stable)}}$ 
shows 
uniqueness of users in the \Dext\ dataset, 
but considering only stable extensions (see more details in Section~\ref{sec:dataset}).
Interestingly, \DextStableUnique\ of users are uniquely identifiable with their stable extensions only and 
the distribution of anonymity set sizes is very similar too.  
This result shows that browser extensions that were added or removed throughout the 9-months-long 
experiment do not influence the result of users' uniqueness.

\begin{figure}[!t]
   \begin{minipage}{0.48\textwidth}
     	\centering
    	\includegraphics[width=1\textwidth]{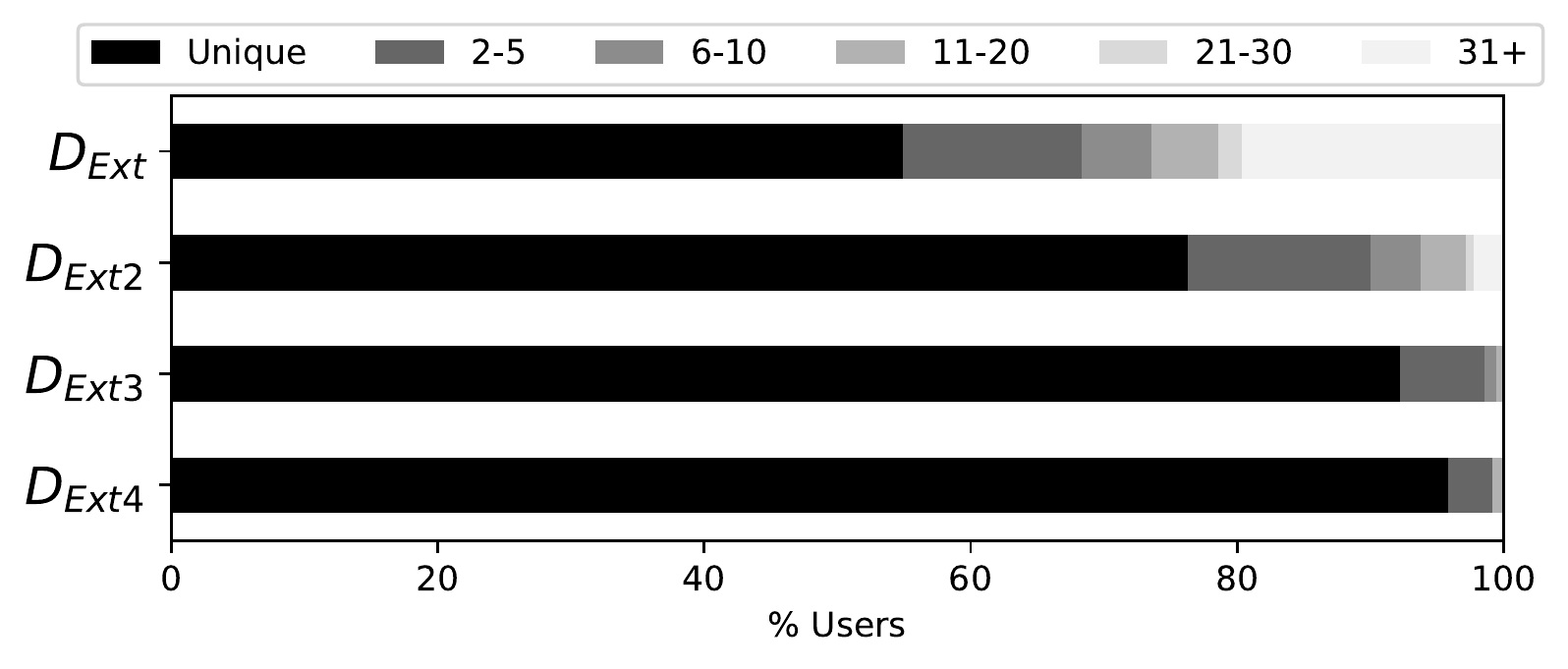}
   \end{minipage}
   \SHORTEN
  \caption{Anonymity sets for users with respect to the number of detected extensions}
  \label{fig:anonymity_sets_nb_extensions_per_user}
\end{figure}

%
%
%
%

{\bf The more extensions you install, the more unique you are.}
In the beginning of this section, we have shown that \DextUnique\ of users are unique among those   
who have at least one extension detected (\Dext\ dataset).
Figure~\ref{fig:anonymity_sets_nb_extensions_per_user} shows how uniquely identifiable users are 
when they have more extensions detected. 
Among users with at least two extensions detected,  \DextAtleastTwoUnique\ are uniquely identifiable. 
This percentage rises quickly to \DextAtleastThreeUnique\ and \DextAtleastFourUnique\ when we consider 
 users with at least three and four extensions detected respectively. 

We made a similar analysis for logins: likewise, the percentage of unique users grows if we consider 
users with a higher number of detected logins. \DlogAtleastFiveUnique\ users with at least 5 logins are uniquely identifiable with their logins only.
This percentage rises to \DlogAtleastEightUnique\ when we detected at least 8 logins.
Intuitively, the more extensions or logins a user has, the more unique he becomes. It is worth mentioning that the subsets of users considered decreases as we increase the number of extensions or logins detected, as shown in Figure~\ref{fig:distribution_number_of_users}.


\begin{figure}[!t]
   \begin{minipage}{0.48\textwidth}
     	\centering
    	\includegraphics[width=1\textwidth]{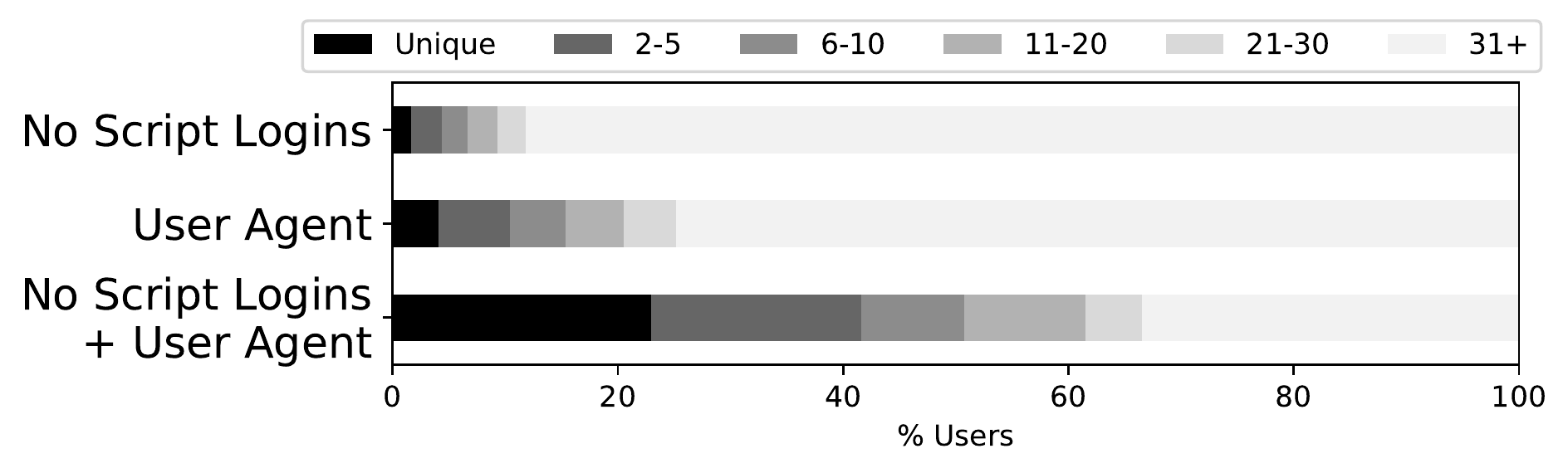}
   \end{minipage}
      \SHORTEN
  \caption{Anonymity sets when JavaScript is disabled}  
  \label{fig:no_js_fingerprintability}
  \SHORTEN
 \end{figure}
 
{\bf Uniqueness if JavaScript is disabled.}
Users might decide to protect themselves from fingerprinting by disabling JavaScript in their browsers.
However even when JavaScript is disabled, 
detection of logins via a CSP violation attack still works. 
Among 60 websites in our experiment, we discovered that 
such an attack works for 18 websites.
Figure~\ref{fig:no_js_fingerprintability} shows 
anonymity sets for \DlogSize\ users of \Dlog\ dataset assuming users have disabled JavaScript. 
By considering only logins detectable with CSP, 1.63\% of users are uniquely identifiable, and 4.10\% are unique based on 
a user agent string that is sent with every request by the browser.
However, 
when we combine the 
user agent string with the list of logins detectable with CSP, 22.98\% of users become uniquely identifiable.

\section{Fingerprinting attacks}
\label{sec:constrained}


According to the uniqueness analysis from Section~\ref{sec:uniqueness}, 
 \DextUnique\ of users that have installed at least one detectable extension are unique; 
 \DlogUnique\ of users are unique  among those who have 
 logged into one or more detectable websites; 
 and \DextAlogUnique\ are unique  among users with at least 
 one extension and one login.
 Therefore, extensions and logins can be used to track users across websites.
In this section we present the threat model,  discuss and evaluate 
two algorithms that optimize fingerprinting based on extensions and logins.

\subsection{Threat model}
The primary attacker is an entity that wishes to uniquely identify a user's browser across websites.
An attacker is recognizing the user by his \emph{browser fingerprint},
a unique set of detected 
browser extensions and Web logins (we call them \emph{attributes}), 
without relying on cookies or other stateful information.
A single JavaScript library that is embedded on a visited webpage can 
check what extensions and Web logins are present in the user's browser. 
By doing so, an attacker is able to uniquely identify the user and track her activities 
across all websites where the attacker's code is present. 
%
%
We assume that an attacker has a dataset of users' fingerprints, either previously 
collected by the attacker or bought from data brokers.

\subsection{How to choose optimal attributes?}
\label{sec:property-to-choose}
The most straightforward way 
to track a user via browser fingerprinting 
is to check all the attributes (browser extensions and logins) of her browser. 
However, testing all \AverageExtensions~extensions 
takes around 30 seconds\footnote{We evaluate performance in Section~\ref{sec:performance}.} 
and thus 
may be unfeasible in practice.
Therefore, the \textit{number of tested attributes} is one of the most important property of fingerprinting attacks -- 
the attack is faster when  fewer attributes are checked.
%
But  testing fewer attributes may lead to worse uniqueness results, because more users will share the same fingerprint.



While it was shown that finding the optimal fingerprint is an NP-hard problem \cite{Guly-Acs-Cast-16-PETS}, finding approximate solutions is neither a trivial task. For example, choosing the most popular attributes worked in the case of tracking based 
on Web history, but this strategy is not necessarily the globally optimal case. 

Following the theoretical results of Guly{\'a}s et al.~\cite{Guly-Acs-Cast-16-PETS}, we consider 
these two strategies: (1) to target a specific user, and thus to select attributes that makes her  unique with high probability
-- called \emph{targeted fingerprinting} algorithm, and (2) to uniquely identify a majority of users in a dataset, 
and thus select the same set of attributes for all users -- we call it \emph{general fingerprinting} algorithm.
Targeted fingerprinting mainly uses popular attributes if they are not detectable 
(e.g., popular extensions are not installed) or unpopular ones if they are detectable. 
General fingerprinting instead, considers attributes that are detectable roughly 
at half of the population (this allows to chose more independent attributes
which makes a fingerprint based on these attributes more unique).

Using the algorithms developed in~\cite{Guly-Acs-Cast-16-PETS}, 
we performed experiments with {general}  and {targeted} fingerprinting. 
Our goal is to 
achieve results close to those in Section~\ref{sec:uniqueness}, 
but by testing a smaller number of attributes\footnote{We reused the implementation of Guly{\'a}s et al., who shared their code \cite{constrainedfingerprinting}.}.

\subsection{Targeted fingerprinting}

{\bf Attack outline.}
The attacker aims to identify a specific user with high probability. 
In order to do this, the attacker needs to have information about the 
targeted user in her dataset of fingerprints. 
The attacker generates a \emph{fingerprint pattern} that consists of a list of 
attributes with a known value, such as
$f_{j}=$ {[AdBlock=yes, LastPass=No, ...]}. 
Notice that a fingerprint pattern contains not only extensions that the user installed, but also 
extensions that are not installed. 
This information also helps to uniquely identify the user.

Let us denote the user database as $D$ of $n$ users and $m$ attributes, 
each row $i$ corresponding to user $U_i$ and each column $j$ corresponding to attribute $A_j$. 
Let the algorithm target user $U_i$. First, we need to find her 
most distinguishing attribute $A_j$, shared among the smallest number of other users.
Let us denote these users as $S_{i,j}$. Then we need to find a second most 
distinguishing property $A_k$ which separates $U_i$ from $S_{i,j}$. Then 
the algorithm continues searching for the most distinguishing
attributes, until the given pattern makes $U_i$ unique 
(or there are no more acceptable choices left).


{\bf Evaluation.}
We applied targeted fingerprinting algorithm~\cite{Guly-Acs-Cast-16-PETS} on our datasets 
$D_{Ext}, D_{Log}$, $D_{Ext} \cap D_{Log}$ and $D_{Ext} \cup D_{Log}$, 
and computed a fingerprint pattern for each user. 
By using these patterns, we have computed the anonymity sets for all 
datasets, that are identical to those 
shown in Figure~\ref{fig:anonymity_sets_datasets}. 
We therefore omit repeating these results in a new figure.

\begin{figure}[t]
\centering
\includegraphics[width=0.5\textwidth]{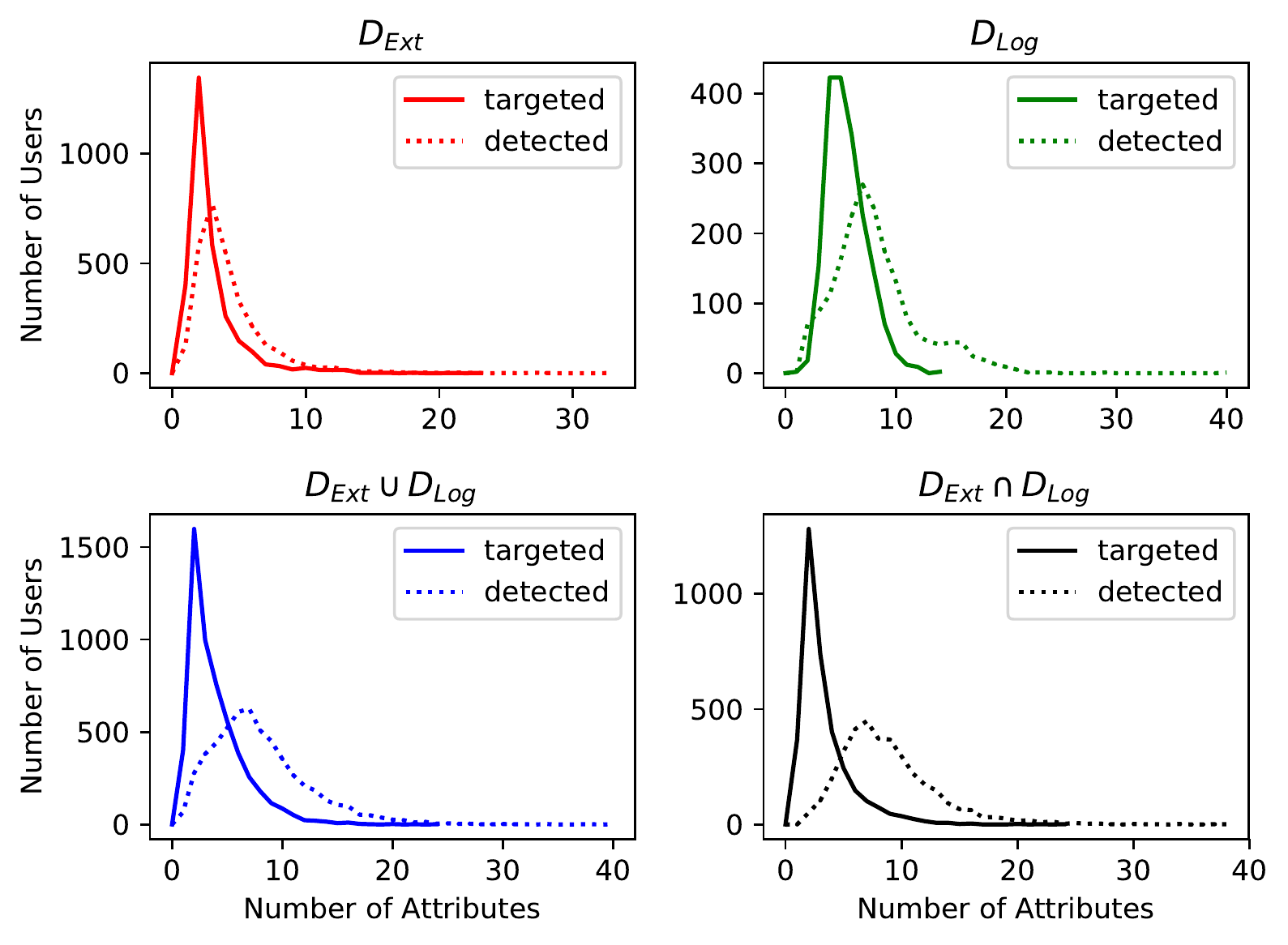}
\caption{Comparison of fingerprint pattern size (targeted) 
and the total number of detected attributes (detected) 
for unique users.}
\label{fig:all_lengths}
\SHORTEN
\end{figure}

For each unique user, the fingerprint pattern contains a smaller number 
of attributes than the number of attributes detected for the user. 
For example, it is enough to test only 2 extensions for a 
user who has installed 4 detectable extensions. Figure \ref{fig:all_lengths} shows 
the distribution of fingerprint pattern sizes of unique users (marked with ``targeted''), 
and compares them to the number of attributes detected for each user. 
The figure clearly shows that fingerprint patterns
are typically smaller than the number of detected attributes users have.

For non-unique users, the size of the fingerprint pattern is often bigger than the 
number of detected attributes the user has.
Let us discuss this on our largest dataset $D_{Ext} \cup D_{Log}$, 
but note that other datasets exhibit the same phenomena. 
For unique users, on average we have 7.94 attributes detected, 
while the average size of fingerprint pattern is 3.94 attributes only. For non-unique users, the average number of detected attributes is 
5.41, while the average size of fingerprint pattern grew up to 30.17. This result is not surprising: 
with less information it is more difficult to distinguish users, and the fingerprint pattern 
may also include negative attributes (i.e., LastPass=No means an extension should not be detected), which can extend the length greatly.

The targeted fingerprint is efficient, as it provides almost maximal uniqueness while reducing the number of attributes. 
However, it cannot be used for new users, because the attacker does not have any background knowledge about them.
To reach a wider usability with a trade-off in the fingerprint pattern size, we also consider general fingerprinting~\cite{Guly-Acs-Cast-16-PETS}.

\subsection{General fingerprinting}
\label{sec:general}

{\bf Attack outline.}
The purpose of this algorithm is to provide a short
list of attributes, called \emph{fingerprint template}. 
If the attributes in a fingerprint template are tested for a 
certain user, she will be uniquely identified with high probability.
Similarly to the example of targeted fingerprinting, we consider the fingerprint template
$F=$ {[AdBlock, LastPass, ...]}, 
that would yield the fingerprint $f^{F}_{j}$=[yes, no, \dots]
for the  user $U_j$.

The algorithm first groups all users into a set $S$. Then it looks for an attribute $A_i$ 
that will separate $S$ into roughly equally sized subsets $S_{1}$ and $S_{2}$ 
if we group users based on their attribute $A_i$. 
In the next round, it looks for another $A_j \neq A_i$ that splits $S_{1}, S_{2}$ further 
into roughly equally sized sets.
This step is repeated until we run out of applicable attributes or the remaining sets could not be sliced further.

\begin{figure*}[t!]
\centering
\subfloat[$D_{Ext}$ - 5,474 users]{
  \includegraphics[width=0.24\textwidth]{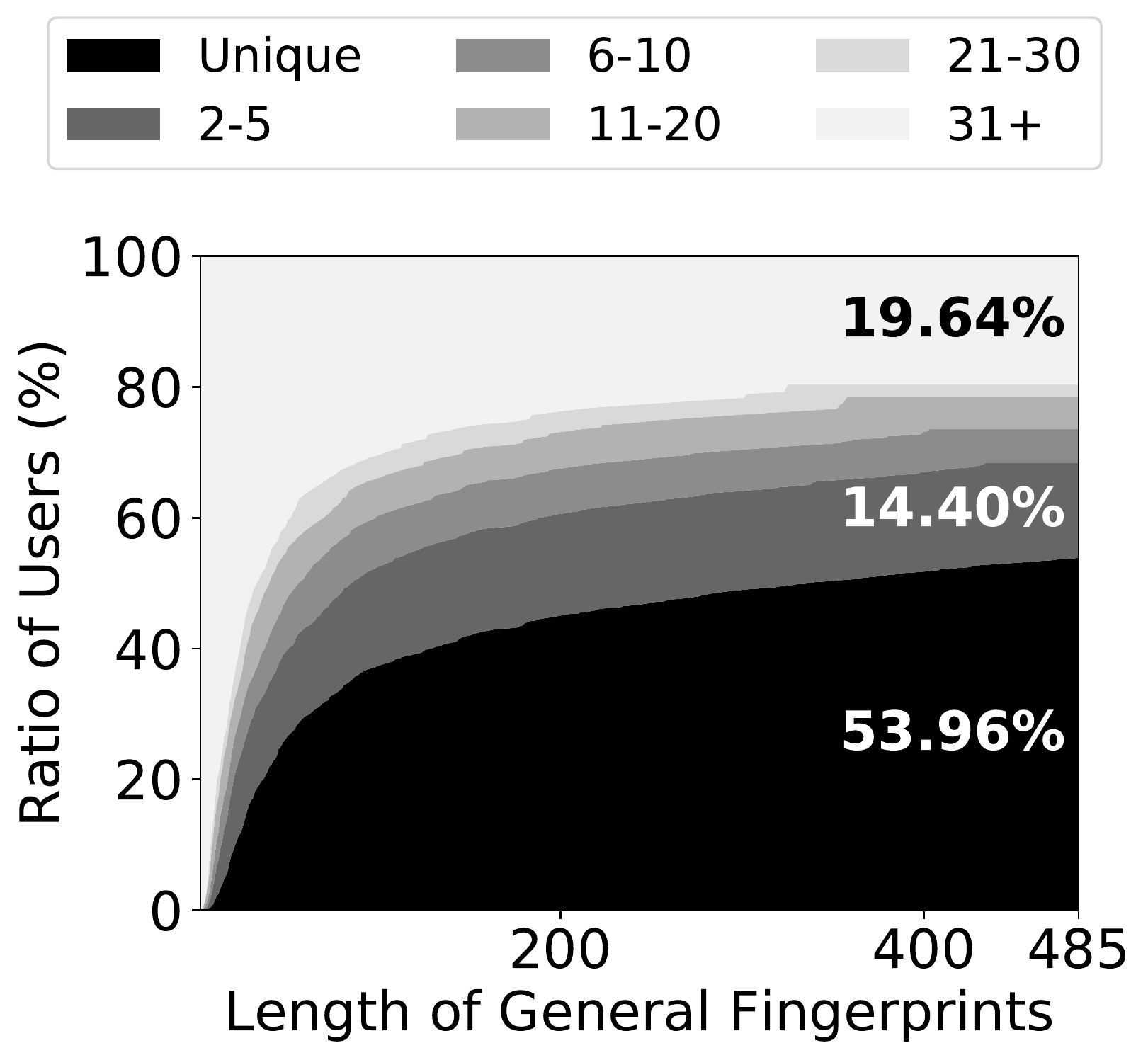}
  \label{fig:general_fingerprint:extension}
}
\subfloat[$D_{Log}$ - 9,492 users]{
  \includegraphics[width=0.24\textwidth]{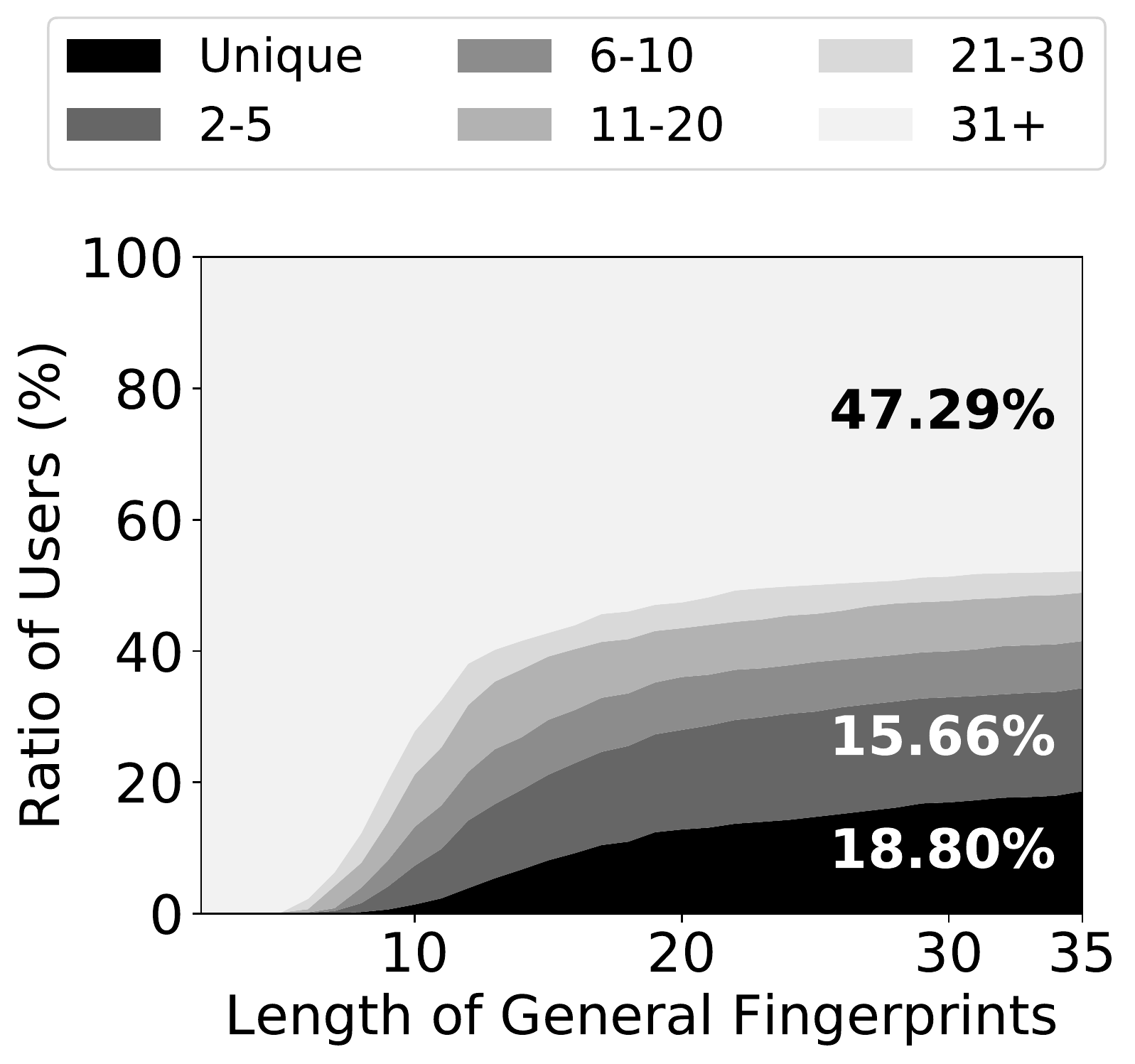}
  \label{fig:general_fingerprint:weblogins}
}
\subfloat[$D_{Ext} \cap D_{Log}$ - 3,919 users]{
  \includegraphics[width=0.24\textwidth]{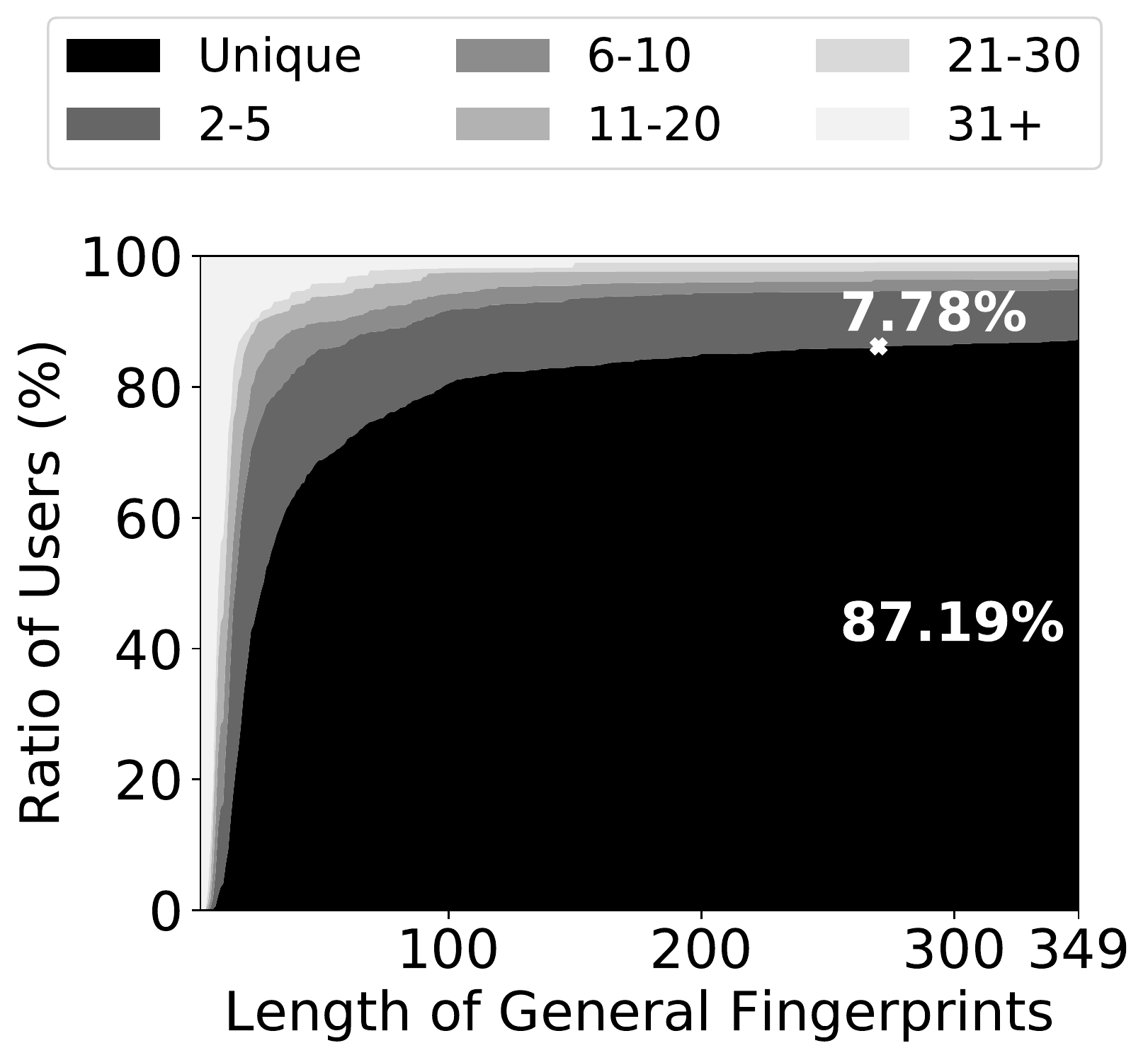}
  \label{fig:general_fingerprint:intersect}
}
\subfloat[$D_{Ext} \cup D_{Log}$ - 11,047 users]{
  \includegraphics[width=0.24\textwidth]{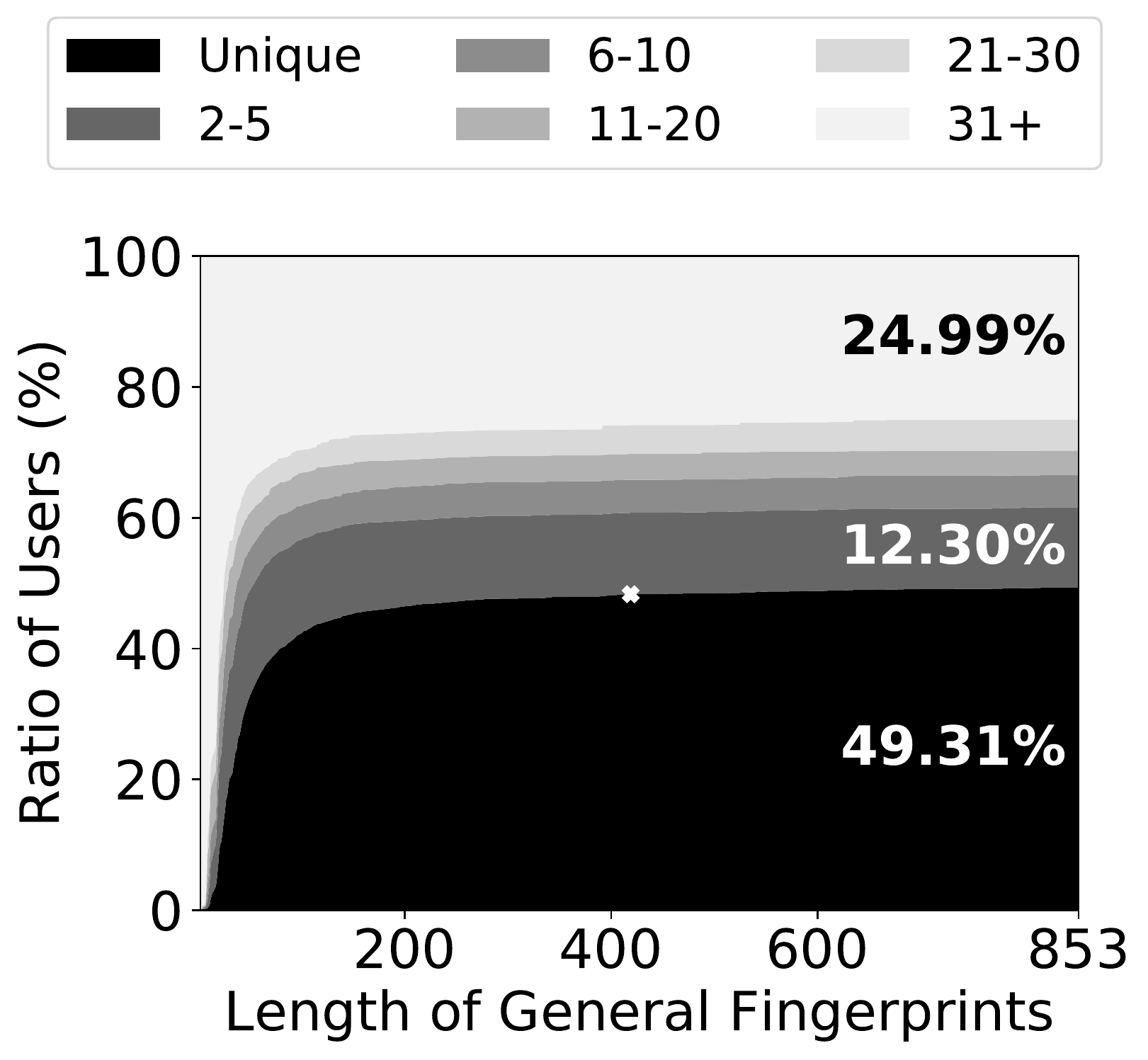}
  \label{fig:general_fingerprint:combined}
}
\caption{Anonymity sets for different numbers of attributes tested by general fingerprinting algorithm.}
\label{fig:general_fingerprint}
\end{figure*}

{\bf Evaluation.}
To apply  general fingerprinting, 
we first measure uniqueness by using all attributes, which will be our target level A. 
Then, we run the algorithm until either it stops by itself (e.g., fingerprint cannot be extended further), 
or we terminate it early when the actual level of uniqueness B is 
less than 1\% from level A. 

Figure \ref{fig:general_fingerprint} shows the anonymity sets for different fingerprint lengths for 
our datasets $D_{Ext}, D_{Log}$, $D_{Ext} \cap D_{Log}$ and $D_{Ext} \cup D_{Log}$, 
generated by the general fingerprinting algorithm. 
For $D_{Ext}$ and $D_{Log}$, the algorithm provided fingerprint templates of  $485$ extensions and $35$ logins.
In these cases the algorithm stopped since no more attributes could be used for achieving better uniqueness -- 
hence the final anonymity sets are very close to those in Figure~\ref{fig:anonymity_sets_datasets}.
In the cases of $D_{Ext} \cap D_{Log}$ and $D_{Ext} \cup D_{Log}$, we observed slow convergence in uniqueness, thus we could stop the algorithm earlier 
(shown as white dots in Figure \ref{fig:general_fingerprint}). 
As a result, for $D_{Ext} \cap D_{Log}$, we can obtain $86.19\%$ of unique users by testing $270$ extensions and logins. 
For $D_{Ext} \cup D_{Log}$, we can obtain $48.31\%$ of unique users by testing $419$ extensions and logins.

We conclude that the general fingerprint can achieve a significant decrease in the fingerprint length while maintaining the level of uniqueness almost at maximum. In the next section we discuss the performance of these results.

For $D_{Ext}$ dataset, general fingerprinting algorithm provides 485 extensions, but we found out that 20 of these extensions were not stable 
(see Figure~\ref{fig:evolution_extensions_wrt_first_month}) and were not present in the last month of our experiment. 
Using all extensions, including unstable ones, can be useful to maintain fingerprint comparability with older data or with 
users having older versions of extensions. However, if we constrain general fingerprinting to stable extensions only, 
we get a fingerprint template of 465 extensions, leading to $50.33\%$ uniqueness -- still very close to the results of baseline uniqueness results, which was $50.35\%$ with stable extensions only.


\section{Implementation and performance}
\label{sec:performance}

In this section we discuss the design choices we made for our experimental website and analyze whether browser extensions and Web logins fingerprinting is efficient enough to be used by tracking companies. 

To collect extensions installed in the user's browser, we first needed to collect the extensions' signatures from the Chrome Web Store. We collected \LastMonthExtensions\ extensions in August 2017, using the code shared by Sj\"{o}sten et al.~\cite{Sjos-etal-17-CODASPY}. 
To detect whether an extension was installed, we tested only one WAR per extension (see more details on WARs in Section~\ref{sec:background}). 
Because the extensions' signatures size was 40Mb and could take a lot of time to load on the client side, we reorganized and compressed them to 600kb.
However, testing all the \LastMonthExtensions\ extensions at once took 11.3--12.5 seconds and was freezing the UI of a Chrome browser. 
To avoid freezing, we split all the extensions in batches of 200 extensions, and testing all the \LastMonthExtensions\ extensions ran in approximately 30s.

Since testing all the extensions takes too long, trackers may not be using this technique in practice. 
Therefore, we measured how much time it takes to apply the optimized fingerprinting algorithms from Section~\ref{sec:constrained}. 
Targeted fingerprinting addresses each user separately, hence the number of tested extensions  differs a lot from user to user. 
General fingerprinting instead provides a generic optimization for all users. Based on our results from Section~\ref{sec:constrained}, 
an attacker can test 485 extensions and obtain the same uniqueness results as with testing all \LastMonthExtensions\ extensions. 
Such testing can be run in 625 milliseconds with the signature file size below 25Kb, which make real-life tracking feasible. For websites with limited traffic volumes, extension detection alone could be used for tracking, or for websites with a higher traffic load, it could contribute supplementary information for fingerprinting.
Regarding targeted fingerprinting the attacker can do even better, as such short patterns can be detected in less than 10 milliseconds.

Compared to extension detection, Web login detection methods depend on more external factors (such as network speed and how fast websites respond), thus they should be used with caution. 
For redirection URL hijacking detection, we observed that the majority of Web logins can be detected in 0.9--2.0 seconds, 
however the timing was much harder to measure for the method based on CSP violation report.
We observed that if the network was overloaded and requests were delayed, then the results of login detection were not reliable; however, it is likely that unreliable results can be easily discarded by checking timings of results (e.g., large delays appearing only in few cases).

Moreover, we found a bug in the CSP reporting implementation in the Chrome browser that makes this kind of detection even more difficult. In fact, without a system reboot for more than a couple of days (we observed that this varies between one day to multiple weeks), the browser stopped sending CSP reports. We reported the issue to Chrome developers, as this bug not only makes CSP-based detection unreliable, but more importantly CSP itself.

\section{The dilemma of privacy extensions}

Various extensions exist that block advertisement content, such as 
AdBlock~\cite{AdBlock}, or block content that tracks users, 
such as Disconnect~\cite{Disconnect}. Such extensions undoubtedly
protect users' privacy, but if they are easily detectable on an arbitrary webpage, 
then they can contribute to users' fingerprint and can be used to track the user across websites. 
In our experiment based on detecting extensions 
via WARs, we could detect four privacy extensions: 
AdBlock~\cite{AdBlock}, Disconnect~\cite{Disconnect}, Ghostery~\cite{Ghostery} and Privacy Badger~\cite{PrivacyBadger}.
The goal of this section is to analyze the tradeoff between the privacy loss (how 
fingerprintable users with such extensions are) 
and the level of protection provided by these extensions.

To understand this tradeoff, 
we computed (i) how unique are the 
users who install privacy extensions; 
(ii) how many third-party cookies are stored in the user's browser 
when a privacy extension is activated (the smaller the number of third-party 
cookies, the better the privacy protection). 
We analyzed four privacy extensions detectable by WARs and 
16 combinations of these extensions: 
AdBlock~\cite{AdBlock}, Disconnect~\cite{Disconnect}, Ghostery~\cite{Ghostery} and Privacy Badger~\cite{PrivacyBadger}.

\begin{figure}[!t]
   \begin{minipage}{0.48\textwidth}
     	\centering
    	\includegraphics[width=3.4in]{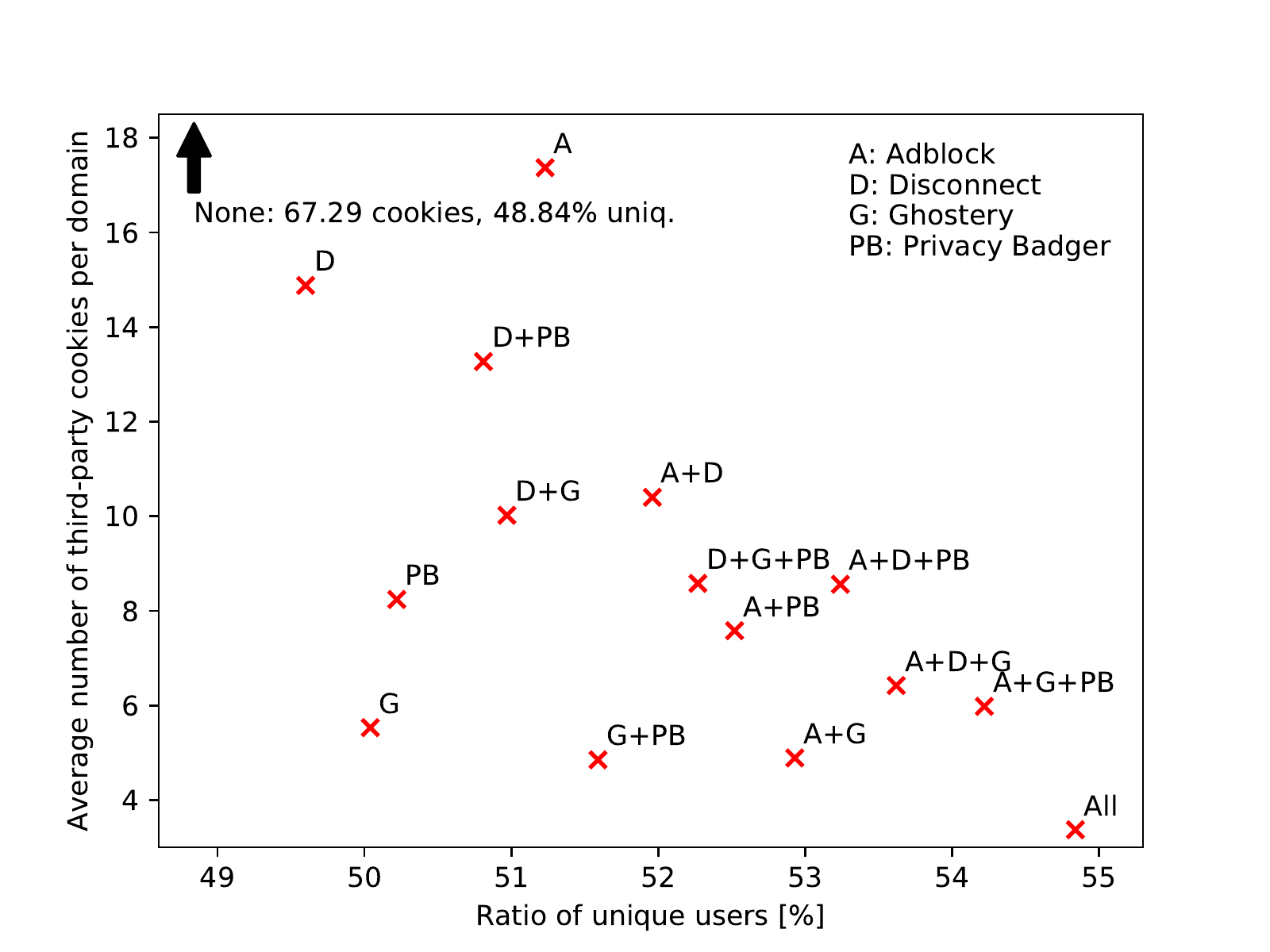}
   \end{minipage}
  \caption{Uniqueness of users vs. number of unblocked third-party cookies}
  \label{fig:uniqueness_vs_cookies}
  \SHORTEN
\end{figure}


First, 
we measured how a
combination of privacy extensions contributes to fingerprinting.
To measure uniqueness of users for a combination of extensions
(i.e., AdBlock+Ghostery), we removed other privacy extensions 
from the \Dext\ dataset\footnote{The total number of users 
does not change since we simply remove certain extensions from the user's 
record in our dataset.}
(i.e., Disconnect and Privacy Badger), and 
then 
evaluated the percentage of unique users for each combination. 

Second, 
we measured how many third-party cookies were set in the browser, even if privacy extensions were enabled. 
We performed an experiment, where  for each combination of extensions,  
we crawled the top 1,000 Alexa domains, visiting homepage and 4 additional pages in each domain\footnote{We have extracted the first 4 links 
on the page that refers to the same domain.}. We kept the browsing profile while visiting pages in the same domain, 
and used a fresh profile when we visited a new domain.
We explicitly activated Ghostery, which is deactivated by default, and trained Privacy Badger on homepages of 1,000 domains before performing our experiment.
%
We collected all the 
third-party cookies that remained in the user's browser for each setting 
and divided it by the number of domains 
crawled. 

Figure \ref{fig:uniqueness_vs_cookies} reports on the average number of cookies that remained 
in the browser for each combination of extensions, and 
the corresponding percentage of unique users.

Similarly to the results of Merzdovnik et al.~\cite{Merz-etal-17-EuroSnP}, 
Ghostery blocks most of the third-party cookies, and the least blocking extension is AdBlock.
Surprisingly, some combinations
such as Disconnect + Ghostery resulted in more third-party cookies being set than for Ghostery alone -- even after double checking the settings, and re-running the measurements, we do not have an explanation for this phenomena. However, as this can have a serious counter-intuitive effect on user privacy, it would be important to investigate this in future work.

More privacy extensions indeed increase user's unicity. 
All of these privacy extensions are also part of the general fingerprint we calculated in Section~\ref{sec:general}.
However, this has little importance in practice. If we ban the general fingerprint algorithm from using 
privacy extensions, it will generate a fingerprint template of 531 (instead of 485) extensions, leading to a uniqueness level of $51.27\%$. While 46 is a significant increase in the number of extensions for fingerprinting, as we have 
seen it already, this would only contribute very little to the overall timing of the attack.

On the other hand, as this experiment revealed, these extensions are also very useful to block 
trackers. We could therefore conclude that using Ghostery is a good trade-off between blocking trackers and 
avoiding extension-based tracking. However, in order to efficiently solve the trade-off dilemma, we 
believe that such functionality should be included by default in all browsers.

\section{Countermeasures}
\label{sec:user_privacy_protection}
We provide recommendations for users who want to be protected from extensions- and logins-based 
fingerprinting. We also provide to developers recommendations to improve browser and extensions architecture in order to 
reduce the privacy risk for their users.

\noindent {\bf Countermeasures for extension detection.} 
Extension detection method based on Web Accessible Resources detects 
\PercentageExtensions\% of Google Chrome extensions,
while for Firefox the number is much smaller:  6.73\% of extensions are detectable by WARs~\cite{Sjos-etal-17-CODASPY}. Firefox gives a good example of browser architecture that makes extensions detection difficult. 
The upcoming Firefox extensions API, WebExtensions, which is compatible with Chrome extensions API~\cite{ChromeExtensionsAPI}, is designed to prevent extensions fingerprinting based on WARs:
each extension is assigned a new random identifier for each user who installs the extension~\cite{WebExtensionsWARs}.
To protect the users, developers of Chrome extensions could avoid 
Web Accessible Resources by hosting them on an external server, however this could lead 
to potential privacy and security problems~\cite{Sjos-etal-17-CODASPY}. 
Developers of the Chrome browser could nonetheless improve the privacy of their users by
adopting the random identifiers for extensions as in WebExtensions API.

Most of the browsers are vulnerable to extension detection, and websites could 
also detect extensions by their behavior~\cite{Star-Niki-17-IEEESP}. Therefore today 
users cannot protect themselves completely, but they  still can minimize the risk by 
using browsers such as Firefox, where a smaller fraction of extensions are detectable.

\noindent {\bf Countermeasures for login.} 
Users may opt for tracker-blocking and adblocking extensions, such as 
Ghostery~\cite{Ghostery}, Disconnect~\cite{Disconnect} or AdBlockPlus~\cite{AdBlockPlus}. 
But these extensions block requests to well-known trackers, 
while Web logins detection sends requests to completely legitimate websites, 
where the user has logged into anyway.
Another option is to install extensions that block  cookies arriving from unknown or 
undesirable domains. These extensions do not protect users for the same reason: 
cookies 
that belong to websites that the user visits (and treated as first-party cookies) 
are the same cookies used for login 
detection (with the only difference that the same cookies are treated as third-party cookies).
For example social websites, such as Facebook or Twitter, use first-party cookies. 
Their social button widgets with third-party cookies may still be allowed by the browser extensions 
in the context of other websites. 
 Therefore, users can protect themselves from Web logins detection, 
 only by \emph{disabling third-party cookies} in their browsers. 
 
Website owners 
could also react  to such potential privacy risk for their users. 
In our case, this would simply mean filtering login URL redirection, and
sanity checking other redirection mechanisms against the CSP-based attack.
Unfortunately, this issue has been known for a while, but 
website owners do not patch 
it because they do not consider this as a serious privacy risk \cite{Linus16}. 

Browser vendors could help avoid login 
detection by blocking third-party 
cookies by default. The new intelligent tracking protection of the Safari 
browser takes a step in the right direction, as it blocks access to third-party 
cookies and deletes them after a while.

\section{Related work}
\label{sec:related}

Since 2006, there have been multiple proposals to detect and enumerate
user's browser extensions~\cite{Grossman06, Koto12, Cattani13, Bryant14}.
Most of them were blog posts that were meant to raise awareness in the
security community, but they did not aim either 
to scientifically evaluate extension detection at large scale,
nor to perform user studies, that could explain
how extensions contribute to browser fingerprinting.
Similarly, there has been an ongoing discussion on Web login detection in the
security community~\cite{Grossman08, Anthony12, Homakov14, Homakov15, Elsobky16, Linus16},
but no quantitative studies have been made until this work.

Sj\"{o}sten et al.~\cite{Sjos-etal-17-CODASPY} provided the first large scale study
on enumerating all free browser extensions that were available to Chrome and Firefox.
While their work lacked the evaluation of user uniqueness or fingerprintability, it disclosed the fact
that 28 of the Alexa top 100k sites already used extensions detection. This result made it clear that
extension detection is more than a theoretical privacy threat, thus deserving further studying.

Starov and Nikiforakis~\cite{Star-Niki-17-IEEESP}
were the first to analyze fingerprintability of browser extensions and evaluating how
unique users are based on their extensions. Differently from our method, they detected
extensions based on the changes extensions make to the webpages. They
examined top 10,000 Chrome extensions and found that 9.2\% of them were detectable
on any website, and 16,6\% made detectable changes on specific domains with 90\%
accuracy.
In contrast, we used Web Accessible Resources~\cite{Sjos-etal-17-CODASPY}
to detect extensions, and analyzed all free Chrome Web Store extensions.
In our measurement period, we observed that 27$-$28\% of all free Chrome extensions were detectable on any
website with 100\% accuracy. While we did not measure this, in the study
of~\cite{Sjos-etal-17-CODASPY} the authors found that 38.96\% of top 10k extensions
in the Chrome Web Store are detectable with WARs.
S{\'{a}}nchez{-}Rola et al.~\cite{Sanc-etal-17-USENIX} detected browser extensions
through a timing side-channel attack, and were able to detect all extensions in Firefox
and Chrome that use access control settings, regardless of the visited site.

Both us and Starov and Nikiforakis analyzed the stability of the proposed detection method.
For a sample of 1,000 extensions, Starov and Nikiforakis concluded that 88\% of
extensions were still detectable after 4 months. In our study, we analyzed
\FirstMonthExtensions\ extensions, and  conclude that 72.4\% of them are
detectable in every month during 9-months period.

To evaluate uniqueness of users based on their browser extensions,
Starov and Nikiforakis have collected installed extensions for 854 users. 
In total, their users had 174 extensions that were fingerprintable.
S{\'{a}}nchez{-}Rola et al.\cite{Sanc-etal-17-USENIX} collected fingerprints
from only 204 users and tested for 2,000 Chrome and Firefox extensions.
In our study, we have \chromeUsers\ Chrome users, for whom we tested \AllExtensions\
extensions, among them \UsedExtensions\ extensions were installed by our users.

Regarding performance, Starov and Nikiforakis~\cite{Star-Niki-17-IEEESP}
reported that to detect 5 extensions, their testing website needed roughly 250 ms.
%
%
S{\'{a}}nchez{-}Rola et al.\cite{Sanc-etal-17-USENIX} are using timing attack, which works
in a very similar way as detecting WARs. 
When querying a non-exist (fake) WAR of an extension, the authors observed a difference in the time the browser takes to respond to the query, depending on whether the extension is installed in the user's browser or not.
The difference is caused
by the access control mechanism of the browser when the concerned extension is installed or not in the browser.
Because of this timing method, S{\'{a}}nchez{-}Rola et al. had to make 10 calls per extension,
while in our work we made only one single call per extension.
We measured that the checking time of non-existing resources and loading existing WARs
are very close to each other (around 1 ms), thus we argue that our approach is significantly faster.

\section{Discussion and future work}

\begin{figure}[!t]
     	\centering
\includegraphics[width=0.45\textwidth]{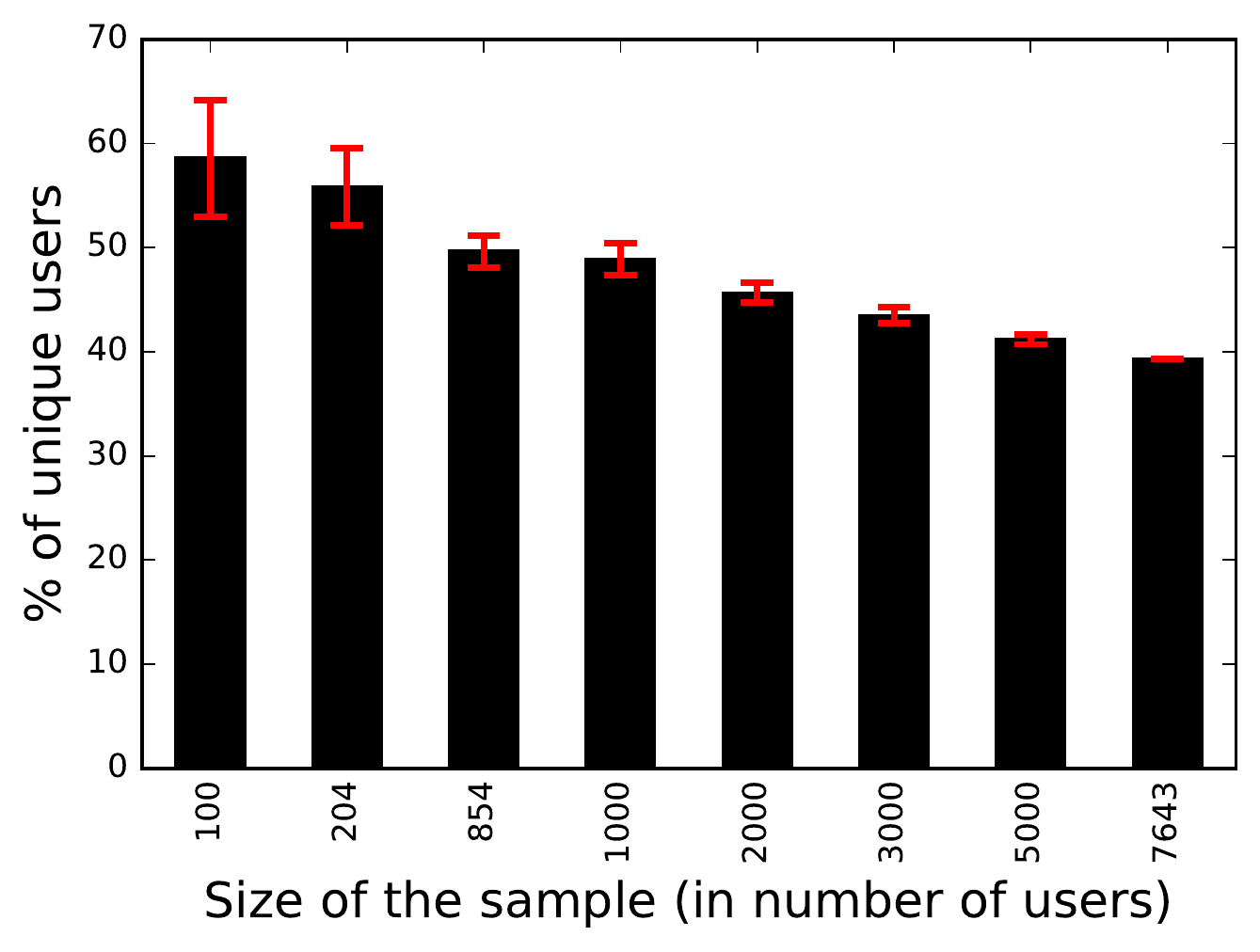}
  \caption{Uniqueness of Chrome users based on their extensions only vs. number of users - 204 is the number of users used in~\cite{Sanc-etal-17-USENIX} and 854 the number of users considered in~\cite{Star-Niki-17-IEEESP}}
  \label{fig:uniqueness_vs_number_of_users}
  \SHORTEN
\end{figure}

{\bf Realistic datasets.}
To compare our study with previous works on fingerprinting by browser extensions, 
we analyzed different random subsets of \chromeUsers\ users, who run Chrome web browser (where 
browser extension detection is possible in our experiment).
Figure~\ref{fig:uniqueness_vs_number_of_users} shows how user uniqueness
based on extensions 
changes with respect to the various subsets of our dataset.
It clearly demonstrates an intuition that 
the smaller the user set is, the smaller is the diversity of users, and the easier 
it is to uniquely identify them. 

Figure~\ref{fig:uniqueness_vs_number_of_users} compares our results to previous 
studies on browser extensions fingerprinting:  we have 
\chromeUsers\ 
Chrome users, while 
previous studies had 204~\cite{Sanc-etal-17-USENIX} and 854~\cite{Star-Niki-17-IEEESP} users, and therefore draw different conclusions 
about uniqueness of users based on browser extensions. 

We reported on the number of unique users in subsets of 
204 and 854 users in Section~\ref{sec:datastats} (see Table~\ref{tab:previous_studies}).
By exploring this comparison, we raise a fundamental question:  
\emph{What is the ``right'' size for the dataset?}

Taking a look at research on standard fingerprinting, in 2010 Eckersley showed that $95\%$ of browsers
were unique based on their properties \cite{Ecke-10-PETS}, which was backed by several papers since 
then~\cite{Boda-etal-11-Nordec, Lape-etal-16-SP}. However, a recent study states that by looking at 
2 million fingerprints in 2018, the authors only found $33.6\%$ of those fingerprints to be unique~\cite{Gome-etal-18-WWW}. 

It is extremely difficult for computer scientists to get access
to such large datasets -- in our experience, we advertised our experiment website through 
all possible channels, including Twitter, Reddit, and press coverage.
We experienced that having larger, high quality datasets is a highly nontrivial research task.
It is important to re-evaluate our results over time while also aiming to obtain larger dataset sizes.

{\bf Stability of fingerprints}
While studying uniqueness based on various behavioural features, 
it is very important to know how stable these features are, as 
the ability to use some of this information as
part of a fingerprint does not solely depend on its anonymity
set of overall entropy, but also on the information stability
(i.e., how frequently it changes over time). 
Vastel et al.~\cite{Vast-etal-18-IEEESP} recently analyzed the evolution 
of fingerprints of 1,905 browsers over two years. They concluded that 
fingerprints' evolution strongly depends on the type of the device (laptop vs mobile)
and how it is used. Overall, they observed that 50\% of browsers 
changed their fingerprints in less than 5 days. 

In our study we did not have enough data to make any claims about the 
stability of the browser extensions and web logins because only few users 
repeated an experiment on our 
website (to be precise, only \UsersMoreThanThreeExperiments\ users out of \FinalUsers\ users 
have made \ThresholdNumerousExperiments\ tests on our website). 
We would expect that browser extensions are more stable than logins since 
users do not seem to change extensions very often, while they may log in and log out 
of various websites during the day. However, studying the stability of extensions and logins 
would require all our users to install a tool (probably a browser extension) in their 
browsers that would monitor the extensions they install and logins they perform. 
This kind of experiment would be even harder to perform at large scale since users 
do not easily trust to install new browser extensions.
In AmIUnique experiment, Laperdrix~\cite{Lape-17-PhDThesis} was trying to measure 
stability of browser fingerprints -- he collected data from 3,528 devices 
over a twenty-month-long experiment. We managed to have \FinalUsers\ users testing 
our website in 9 months. This shows that users have more trust in testing 
their browser on a website than installing new extensions.

We therefore keep the study of fingerprints stability for future 
work and raise an important question  in privacy measurement community: 
\emph{How can we ensure a large scale coverage of users for our privacy 
measurement experiments?}

\section{Conclusion}

This paper reports on a large-scale study of a new
form of browser fingerprinting technique based on browser extensions
and website logins. The results show that 
\ExtUnique\% of users are unique because of the extensions they install (\DextUnique\ of users that have installed at least one detectable extension
are unique); \LogUnique\% of users are unique because of the websites they are logged into (\DlogUnique\ are unique among those who have logged into one or more detectable websites); and \ExtOLogUnique\% of users are unique when combining their detected extensions and logins (\DextAlogUnique\ are unique among users with at least one extension and one login).
It also shows that the
fingerprinting techniques can be optimized and performed in 625 ms.

This paper illustrates, one more time, that user anonymity is very challenging 
on the Web.
Users are unique in many different ways in the real life and on the Web.
For example, it has been shown that users are unique in the way they browse
the Web, 
the way they move their mouse or by the applications they install on their device~\cite{Haye-14-CS}.
This paper shows that users are also unique in the way they configure and augment their browser, and by the sites
they connect to. Unfortunately, although uniqueness is valuable in society because it increases 
diversity, it can be misused by malicious websites to fingerprint users and can therefore hurt privacy.

Another important contribution of this paper is the definition and the study of the trade-off that exists when a user 
decides to install a ``privacy'' extension, for example, an extension that blocks trackers.
This paper shows that some of these extensions increase user's unicity and can therefore contribute to
fingerprinting, which is counter-productive.
We argue that these ``privacy'' extensions are very useful, but they should be included by default in all browsers. ``Privacy by default'', as
 advocated by the new EU privacy regulation, should be enforced to improve privacy of all Web users.

\section*{Acknowledgment}

First of all, we would like to thank the valuable comments and suggestions of the anonymous reviewers of our paper. We are grateful for Alexander Sj\"osten, Steven Van Acker, Andrei Sabelfeld, who shared their code and signature database for Chrome browser extension detection \cite{Sjos-etal-17-CODASPY}, and also for Robin Linus, who allowed us to build on his script on social media presence detection. We thank Imane Fouad for supporting our work.

\bibliographystyle{abbrv}
\bibliography{bibfile.bib}

\end{document}